\documentclass[%
aip,
rsi,%
amsmath,amssymb,
preprint,%
]{revtex4-1}
\usepackage{nicefrac}
\usepackage{units}
\usepackage{amsmath}
\usepackage{appendix}
\usepackage{graphicx}
\usepackage[top=1.5in, bottom=1.5in, left=1in, right=1in]{geometry}
\usepackage{amsfonts}
\usepackage{amssymb}

\begin{document}

\title{Random walks with thermalizing collisions in bounded regions;  \ physical applications valid from the ballistic to diffusive regimes.}
\author{C.M. Swank}
\affiliation{Division of Physics, Math and Astronomy, California Institute of Technology, Pasadena, CA}
\author{ A.K. Petukhov}
\affiliation{Institut Laue-Langevin, BP156, 38042 Grenoble Cedex 9, France} 
\author{ R. Golub}
\affiliation{Physics Department, North Carolina State University, Raleigh, NC 27695}

%\tableofcontents
 \begin{abstract}
 	The behavior of a spin undergoing Larmor precession in the presence of
 	fluctuating fields is of interest to workers in many fields. The fluctuating
 	fields cause frequency shifts and relaxation which are related to their power
 	spectrum, which can be determined by taking the Fourier transform of the
 	auto-correlation functions of the field fluctuations. Recently we have shown
 	how to calculate these correlation functions for all values of mean free path
 	(ballistic to diffusive motion) in finite bounded regions, using the model of
 	persistent continuous time random walks (CTRW) for particles subject to
 	scattering by fixed (frozen) scattering centers so that the speed of the
 	moving particles is not changed by the collisions. In this work we show how
 	scattering with energy exchange from an ensemble of \ scatterers in thermal
 	equilibrium can be incorporated into the CTRW. We present results for 1,2 and
 	3 dimensions. The results agree for all these cases contrary to the previously
 	studied 'frozen' models. Our results for the velocity autocorrelation function
 	show a long time tail $\left(  \sim t^{-1/2}\right)  $, which we also obtain
 	from conventional diffusion theory, with the same power, independent of dimensionality.
 	
 	Our results are valid for any Markovian scattering kernel as well as any
 	kernel based on a scattering cross section $\sim1/v.$
 	
 \end{abstract}

\maketitle

\section{Introduction}

The dynamics of a system of spins moving under the influence of static and
time-varying magnetic fields is a subject of wide ranging scientific and
technical interest. Both randomly fluctuating fields produced by a thermal
reservoir, and fluctuations seen by particles undergoing stochastic
trajectories in inhomogeneous fields have been the subject of intense study
over many decades. It was Bloembergen, Purcell and Pound
\cite{Bloembergen1948} who first showed, using physical arguments based on
Fermi's golden rule, that the relaxation rate is determined by the power
spectrum of the fluctuating fields evaluated at the Larmor frequency. In
general power spectra of fluctuating quantities are given by the Fourier
transform of the auto-correlation of the fluctuating variable. A\ short
history of the development of the field can be found in the introductions to
\cite{Golub2010c, pignol2015}.

One of the applications of these techniques is to the next generation searches
for a particle electric dipole moment(EDM) which require measurements of spin
dynamics in uniform magnetic fields with nanohertz precision. Furthermore,
searches for fundamental forces beyond the standard model require similar
accuracy in the measurement of Longitudinal and Transverse relaxation ($T_{1}$
and $T_{2}$) expected to be produced by the hypothesized interaction. The
desire for such accurate predictions inspires the search for models of
particle trajectories in the case of particles moving in inhomogeneous fields.

Redfield \cite{redfield}, as elucidated by Slichter \cite{slichter}, and
McGregor \cite{mcgregor} have given formal derivations of the relation between
relaxation and the autocorrelation functions of the fields and the method was
applied to a 'false edm' systematic error effecting searches for time reversal
and parity violating non-zero particle electric dipole moments
\cite{pendlebury04,Lam05}. General methods for obtaining auto-correlation
functions for fluctuations produced by particles diffusing in inhomogeneous
fields with arbitrary spatial variation have been given by \cite{petukhov} and
\cite{clayton}.

This work has been extended by \cite{swank} to the case of arbitrary field
variation and all values of scattering mean free path (from ballistic to
diffusive motion) in restricted geometries. The method used was based on the
persistent continuous time random walk model of Masoliver et al,
\cite{masoliver}, who solved a transport equation for the Laplace-Fourier
transform of the conditional probability $P\left(  \overrightarrow
{r},t\right)  $ (the probability that a particle located at$\overrightarrow
{r}=0$ at $t=0$, will be found at position $\overrightarrow{r}$ at time $t$
(sometimes called a propagator), for the case of an infinite domain. The model
assumed a collection of fixed scattering centers ('frozen' environment) so
that the velocity of the particles was unchanged by the scattering events, and
was valid for all values of the mean free path. The authors of \cite{swank}
applied the results of Masoliver et al, \cite{masoliver} to 2 and 3
dimensional regions bounded by rectangles, by using the method of images and
used the resulting conditional probabilities to calculate a number of spectra
of autocorrelation functions relevant to relaxation and frequency shifts in a
range of problems. See \cite{pignol2015} for an overview of the relation
between correlation functions of fluctuating fields and physical phenomena.

In the present work we apply the technique of Swank et al \cite{swank} to the
case of Markovian scattering in which each collision completely re-thermalizes
the scattered particles. The method applies equally to the case when the total
inelastic scattering cross section $\sim1/v$ with $v$ the particle velocity.
We find that the results differ somewhat from those obtained by averaging the
results for the frozen environment over the velocity distribution of the
scattered particles and that the results for 1, 2 and 3 dimensions are
identical when averaged over a Maxwell distribution. Further we show that for stochastic bounded motion the
velocity autocorrelation functions have long time tails proportional to
$t^{-1/2}$, in all cases where diffusion theory is valid, in agreement with
the one dimensional treatment in reference \cite{oppenheim1964}. Other studies
have shown that in non-bounded systems long-range hydrodynamic forces lead to different results
$\left(  \sim t^{-d/2}\right)  $ where $d$ is the number of dimensions of the
system. We present the results of applying our method to the physically
interesting problems of the false edm systematic error in searches for
particle electric dipole moments and to calculating the position and velocity
autocorrelation functions for particles confined to a bounded region, which
determine frequency shifts and relaxation rates in nmr, \cite{pignol2015}.

\section{The Model}

\subsection{Preliminaries: Persistent continuous time random walk in the
frozen environment.}

In this section we review the solution for the spectrum of the probability
density of a persistent continuous time random walk (CTRW) as presented in the
work of Weiss and co-workers \cite{masoliver}. The particles are assumed to
travel ballistically with fixed velocity $v$ between scattering events. The
time between scattering events is governed by a distribution $\psi(t)$, such
that the probability to scatter within a time segment $dt$ is given by
$\psi(t)dt$, and the probability to reach $t$ without scattering is given by
$\Psi(t)=\int_{t}^{\infty}\psi(t)dt$. The conditional probability
$p(\mathbf{x},t)$, is calculated, as well as a scattering density
$\rho(\mathbf{x},t)$, the probability of scattering at $\mathbf{x}$ and time
$t$. A recursive equation that completely describes the CTRW is formed for the
two densities,

\begin{align}
\rho(\mathbf{x},t,v,\Omega)=  &  f(\mathbf{x},t,v)\alpha(\Omega)\psi\left(
t\right) \nonumber\\
&  +\int\int\int d^{3}\mathbf{x}^{\prime}dt^{\prime}d\Omega^{\prime
}f(\mathbf{x-x}^{\prime},t-t^{\prime},v)\beta(\Omega|\Omega^{\prime}%
)\psi\left(  t-t^{\prime}\right)  \int dv^{\prime}\rho(\mathbf{x}^{\prime
},t^{\prime},v^{\prime}),\nonumber\\
p (\mathbf{x},t,v,\Omega)=  &  f(\mathbf{x},t,v)\alpha(\Omega)\Psi\left(
t\right) \\
&  +\int\int\int d^{3}\mathbf{x}^{\prime}dt^{\prime}d\Omega^{\prime
}f(\mathbf{x-x}^{\prime},t-t^{\prime},v)\beta(\Omega|\Omega^{\prime}%
)\Psi\left(  t-t^{\prime}\right)  \int dv^{\prime}\rho(\mathbf{x}^{\prime
},t^{\prime},v^{\prime}). \label{eq:mas01}%
\end{align}

In three dimensions the angular coordinates denoted by $\Omega=\{\theta
,\phi\}$ has element $d\Omega=\mathrm{sin}\theta d\theta d\phi$.
$\alpha(\Omega)$ is the initial angular density, while $\beta(\Omega
|\Omega^{\prime})$ is the conditional angular density having scattered from a
previous angle $\Omega^{\prime}$ also known as the scattering kernel. We will
assume the initial angular density to be isotropic and the angular conditional
density to be isotropic and Markovian. Therefore in three dimensions we have,
\[
\alpha(\Omega)=\beta(\Omega|\Omega^{\prime})=\frac{1}{4\pi}.
\]
The scattering time density will be assumed to follow a simple Poisson
distribution:
\begin{align}
\psi(t)=\frac{1}{\tau_{c}}e^{-\frac{t}{\tau_{c}}}.
\end{align}
where $\tau_{c}$ is the average collision time. The spectrum of the
conditional density is found by applying the Laplace-Fourier Transform to
equations (\ref{eq:mas01}$)$ \cite{masoliver}:%

\begin{align}
p(\mathbf{q},s)=\tau_{c}\frac{\mathrm{arctan}\left(  \frac{qv\tau_{c}}%
{1+s\tau_{c}}\right)  }{qv\tau_{c}-\mathrm{arctan}\left(  \frac{qv\tau_{c}%
}{1+s\tau_{c}}\right)  }.\label{eq:frozen3Dpdf}%
\end{align}
In our previous work \cite{swank}, we extended the free space solution shown
in equation~(\ref{eq:frozen3Dpdf}) (for the 3D case) to the restricted domain
in 1, 2 and 3 dimensions. In the present work we allow the velocity to change
upon a scattering event, changing the model from a ''frozen'' model with fixed
speed, to one that allows momentum transfer. The approach is similar to that
shown in reference \cite{zab}. We will see that the result differs from simply
averaging the single velocity conditional density over velocity and that the
results for three dimensions are identical to the results for lower
dimensions, and a method for predicting three dimensional results from a one
dimensional model (in Cartesian coordinates) is obtained. Results for the
position and velocity autocorrelation functions and applications to the
bounded domain are presented.

\subsection{From the frozen environment to thermalization with momentum transfer. \label{ctrwt}}

In the following we present our model of a CTRW with thermalization that we
refer to as CTRWT. We start from the approach described
above,~\cite{masoliver}. A change in velocity upon a gas scattering can be
accounted for by including a probability distribution for the outgoing
velocity after a scattering event. The treatment of $v~$is identical to that
of reference~\cite{masoliver} for the angular density, except now we allow the
vector velocity $\mathbf{v}$ to change. Therefore we extend functions
$\alpha(\Omega)$ and $\beta(\Omega|\Omega^{\prime})$ $\rightarrow
\alpha(\mathbf{v})$ and $\beta(\mathbf{v}|\mathbf{v}^{\prime})$. Now
$\alpha(\mathbf{v})$ is the initial probability distribution of velocities
with angle $\Omega$ and speed $v$ and $\beta(\mathbf{v}|\mathbf{v}^{\prime}%
)$is the probability of scattering into angle $\Omega$ and speed $v$ with
incoming angle $\Omega^{\prime}$ and speed $v^{\prime}$ prior to the
collision,
\begin{equation}
\rho(\mathbf{x},t,\mathbf{v})=f(\mathbf{x},t,\mathbf{v})\alpha(\mathbf{v}%
)\psi\left(  t\right)  +\int d^{3}\mathbf{x}^{\prime}dt^{\prime}%
f(\mathbf{x-x}^{\prime},t-t^{\prime},\mathbf{v})\psi\left(  t-t^{\prime
}\right)  \int d^{3}\mathbf{v}^{\prime}\beta(\mathbf{v}|\mathbf{v}^{\prime
})\rho(\mathbf{x}^{\prime},t^{\prime},\mathbf{v}^{\prime}). \label{aaa}%
\end{equation}
Where $f(\mathbf{x},t,\mathbf{v})$ is given by,
\begin{align}
f(\mathbf{x},t,\mathbf{v})=\delta^{\left(  N\right)  }\left(  \mathbf{x}%
-\mathbf{v}t\right),
\end{align}
for N dimensions. This is similar to the formulation in $\cite{zab}$, where
they derive the spectrum of the conditional density and correlation functions
in one dimension for arbitrary scattering time densities.

\ \ Now the scattering density,%
\begin{equation}
\rho(\mathbf{x}^{\prime},t^{\prime},\mathbf{v}^{\prime})=N_{s}\sigma
_{tot}\left(  v^{\prime}\right)  v^{\prime}n\left(  \mathbf{x}^{\prime
},t^{\prime},\mathbf{v}^{\prime}\right),
\end{equation}

where $N_{s}$ is the number of scatterers per unit volume, $\sigma
_{tot}\left(  v^{\prime}\right)  $ is the total inelastic scattering cross
section and $n\left(  \mathbf{x}^{\prime},t^{\prime},\mathbf{v}^{\prime
}\right)  $ is the density of particles with velocity $v^{\prime}$ \ at
$\left(  \mathbf{x}^{\prime},t^{\prime}\right)  .$ The double differential
cross section $\sigma\left(  \mathbf{v}^{\prime}\rightarrow\mathbf{v}\right)
=\beta(\mathbf{v}|\mathbf{v}^{\prime})\sigma_{tot}\left(  v^{\prime}\right)  .$

For a system in thermal equilibrium:
\begin{equation}
n(\mathbf{x}^{\prime},t^{\prime},\mathbf{v}^{\prime})=\alpha\left(
\mathbf{v}^{\prime}\right)  n(\mathbf{x}^{\prime},t^{\prime}).
\end{equation}
For the common case $\sigma_{tot}\left(  v^{\prime}\right)  \propto
1/v^{\prime},$ we can write,
\begin{equation}
\rho(\mathbf{x}^{\prime},t^{\prime},\mathbf{v}^{\prime})=\alpha\left(
\mathbf{v}^{\prime}\right)  \rho(\mathbf{x}^{\prime},t^{\prime}),
\label{eq:rhovalphav}%
\end{equation}

where $\rho(x^{\prime},t^{\prime})=N_{s}\sigma_{tot}\left(  v^{\prime}\right)
v^{\prime}n\left(  \mathbf{x}^{\prime},t^{\prime}\right)  $ \ is then
independent of \ $v^{\prime}.$

Thus the second term in (\ref{aaa}) becomes,%
\begin{align}
&  \int d^{3}\mathbf{x}^{\prime}dt^{\prime}f(\mathbf{\ x-x}^{\prime
},t-t^{\prime},\mathbf{v})\psi\left(  t-t^{\prime}\right)  \int d^{3}%
\mathbf{v}^{\prime}\beta(\mathbf{v}|\mathbf{v}^{\prime})\alpha\left(
\mathbf{v}^{\prime}\right)  \rho(\mathbf{x}^{\prime},t^{\prime})\nonumber\\
&  =\alpha\left(  \mathbf{v}\right)  \int d^{3}\mathbf{x}^{\prime}dt^{\prime
}f(\mathbf{\ x-x}^{\prime},t-t^{\prime},\mathbf{v})\psi\left(  t-t^{\prime
}\right)  \rho(\mathbf{x}^{\prime},t^{\prime}), \label{bgbg}%
\end{align}
making use of the property,
\begin{equation}
\int d^{3}\mathbf{v}^{\prime}\beta(\mathbf{v}|\mathbf{v}^{\prime}%
)\alpha\left(  \mathbf{v}^{\prime}\right)  =\alpha\left(  \mathbf{v}\right),
\label{eq:MBcond}%
\end{equation}
which must be satisfied by any physically allowable kernel that produces a
Maxwellian steady state. Thus our method is valid for a variety of
\ experimentally relevant collision kernels such as the cusp kernels
introduced in \cite{McGuyer2012}. For a Markovian thermalization process
$\beta\left(  \mathbf{v|v}^{\prime}\right)  =\alpha\left(  \mathbf{v}\right)
$ independent of $\mathbf{v}^{\prime}$ and (\ref{bgbg}) follows directly from
equation (\ref{aaa}).

With this included our transport equations become,%

\begin{align}
\rho(\mathbf{x},t,\mathbf{v})  &  =\alpha(\mathbf{v})f(\mathbf{x}%
,t,\mathbf{v})\psi\left(  t\right)  +~\alpha(\mathbf{v})\int d^{3}%
\mathbf{x}^{\prime}dt^{\prime}f(\mathbf{x}-\mathbf{x}^{\prime},t-t^{\prime
},\mathbf{v})\psi\left(  t-t^{\prime}\right)  \rho(\mathbf{x}^{\prime
},t^{\prime}),\label{bg2}\\
p(\mathbf{x},t,\mathbf{v})  &  =\alpha(\mathbf{v})f(\mathbf{x},t,\mathbf{v}%
)\Psi\left(  t\right)  +~\alpha(\mathbf{v})\int d^{3}\mathbf{x}^{\prime
}dt^{\prime}f(\mathbf{x}-\mathbf{x}^{\prime},t-t^{\prime},\mathbf{v}%
)\Psi\left(  t-t^{\prime}\right)  \rho(\mathbf{x}^{\prime},t^{\prime}).
\label{bg3}%
\end{align}

The remarkable property of our model (\ref{bg2}, \ref{bg3}) is that it is
independent of the form of the scattering kernel as as long as
(\ref{eq:MBcond}) is satisfied.

Since we are mainly interested in finding the velocity averaged probability,
$p(\mathbf{q,}s\mathbf{),~}$We introduce the velocity integrated quantities,%

\begin{align}
p(\mathbf{x},t)  &  =\int p(\mathbf{x},t,\mathbf{v})d^{3}\mathbf{v,}\\
\rho(\mathbf{x},t)  &  =\int\rho(\mathbf{x},t,\mathbf{v})d^{3}\mathbf{v.}
\label{aaa1}%
\end{align}

The first term in equation~(\ref{bg2}) represents all of the particles at
$\left(  \mathbf{x},t\right)  $ that have not scattered. The second term, a
convolution of the $f$ propagator and scattering density $\rho$ represents
particles that have scattered at $\left(  \mathbf{x}^{\prime},t^{\prime
}\right)  $ and traveled to $(\mathbf{x},t)$ without collision. From here they
can make another collision (\ref{bg2}) or continue on the same path without
scattering, but they contribute to the particle density at $\left(
\mathbf{x},t\right)  $ (\ref{bg3}).

We will take advantage of the convolution theorem of the Fourier-Laplace
transform to solve for the spectrum, $p(\mathbf{q},s).$ Setting,%

\begin{align}
g(\mathbf{x-x}^{\prime},t-t^{\prime},\mathbf{v})  &  =f(\mathbf{x}%
-\mathbf{x}^{\prime},t-t^{\prime},\mathbf{v})\psi\left(  t-t^{\prime}\right),
\\
G(\mathbf{x-x}^{\prime},t-t^{\prime},\mathbf{v})  &  =f(\mathbf{x}%
-\mathbf{x}^{\prime},t-t^{\prime},\mathbf{v})\Psi\left(  t-t^{\prime}\right),
\nonumber
\end{align}
we have from (\ref{bg2}),
\begin{align}
\rho(\mathbf{q,}s,\mathbf{v)}  &  =\alpha(\mathbf{v})g(\mathbf{q}%
,s,\mathbf{v})+\rho(\mathbf{q},s)\alpha(\mathbf{v})g(\mathbf{q},s,\mathbf{v}%
),\\
&  =\alpha(\mathbf{v})g(\mathbf{q},s,\mathbf{v)}\left(  1+\rho(\mathbf{q}%
,s)\right), \label{BG!}\\
\rho(\mathbf{q,}s\mathbf{)}  &  \mathbf{=}\left(  1+\rho(\mathbf{q},s)\right)
\int\alpha(\mathbf{v})g(\mathbf{q},s,\mathbf{v)d}^{3}\mathbf{v},\\
&  =\frac{\int\alpha(\mathbf{v})g(\mathbf{q},s,\mathbf{v)d}^{3}\mathbf{v}%
}{1-\int\alpha(\mathbf{v})g(\mathbf{q},s,\mathbf{v)d}^{3}\mathbf{v}},\\
p(\mathbf{q,}s,\mathbf{v)}  &  =\alpha(\mathbf{v})G(\mathbf{q},s,\mathbf{v}%
)+\rho(\mathbf{q},s)\alpha(\mathbf{v})G(\mathbf{q},s,\mathbf{v}),\nonumber\\
&  =\alpha(\mathbf{v})G(\mathbf{q},s,\mathbf{v})\left(  1+\rho(\mathbf{q}%
,s)\right), \\
&  =\frac{\alpha(\mathbf{v})G(\mathbf{q},s,\mathbf{v})}{1-\int\alpha
(\mathbf{v})g(\mathbf{q},s,\mathbf{v)d}^{3}\mathbf{v}},%
\end{align}
so that,
\begin{equation}
p(\mathbf{q},s)=\frac{\int\alpha(\mathbf{v})G(\mathbf{q},s,\mathbf{v})d^{3}\mathbf{v}%
}{1-\int\alpha(\mathbf{v})g(\mathbf{q},s,\mathbf{v})d^{3}\mathbf{v}}.
\label{11}%
\end{equation}

For gas collisions that randomize velocity after each collision the correct
conditional probability density, $p,$ is not a direct velocity average of the
single velocity $p$, but a function of the velocity average of the individual
propagators of $G$ and $g$.

The collision time and the probability of scattering $\psi(t)$ remain the same
for all three dimensions,
\[
\psi(t)=\frac{1}{\tau_{c}}e^{-\frac{1}{\tau_{c}}t},
\]
where $\frac{1}{\tau_{c}}$ is the rate of gas collisions. The probability of
not making a scattering in time $t$ is given by the integration over the
scattering rate,%

\begin{align}
\Psi(t)=\int_{t}^{\infty}\psi(t)dt=e^{-\frac{t}{\tau_{c}}}.
\end{align}

We define:
\begin{align}
F_{N}(\mathbf{x},t)=\int\alpha_{N}\left(  \mathbf{v}\right)  G_{N}%
(\mathbf{x},t,\mathbf{v})d^{N}\mathbf{v},%
\end{align}
where N represents the number of dimensions in the random walk. We now find 
the Fourier-Laplace transform  of $F_{N}(\mathbf{x},t)$,
\begin{equation}
F_{N}(\mathbf{q},s)=\int_{0}^{\infty}dt\int\alpha_{N}\left(  \mathbf{v}%
\right)  \delta^{(N)}\left(  \mathbf{x}-\mathbf{v}t\right)  e^{-\frac{t}%
{\tau_{c}}-i\mathbf{q\cdot x}-st}d^{N}\mathbf{x}d^{N}\mathbf{v}.\label{eq:Fqs1}%
\end{equation}
We will use the Maxwellian velocity distribution:
\begin{align}
\alpha_{N}\left(  \mathbf{v}\right)  =\prod\limits_{i=1}^{N}\left(  \frac
{1}{2}\sqrt{\frac{2m}{\pi kT}}\right)  e^{-\frac{m}{2kT}v_{i}^{2}}.%
\end{align}
Substituting this into equation (\ref{eq:Fqs1}),
\begin{align}
F_{N}(\mathbf{q},s)=\int_{0}^{\infty}dt\int\left[  \prod\limits_{i=1}%
^{N}\left(  \frac{1}{2}\sqrt{\frac{2m}{\pi kT}}\right)  e^{-\frac{m}{2kT}%
v_{i}^{2}}\right]  \delta^{(N)}\left(  \mathbf{x}-\mathbf{v}t\right)
e^{-\frac{t}{\tau_{c}}-i\mathbf{q\cdot x}-st}d^{N}\mathbf{x}d^{N}\mathbf{v},%
\end{align}
and integration over position gives
\begin{align}
F_{N}(\mathbf{q},s)=\int_{0}^{\infty}dte^{-\left(  s+\frac{1}{\tau_{c}%
}\right)  t}\prod\limits_{i=1}^{N}\left(  \frac{1}{2}\sqrt{\frac{2m}{\pi kT}%
}\right)  \int e^{-\frac{m}{2kT}v_{i}^{2}-iq_{i}v_{i}t}dv_{i},%
\end{align}
integration over $v_{i}$ then gives,
\begin{align}
F_{N}(\mathbf{q},s) &  =\int_{0}^{\infty}dte^{-\left(  s+\frac{1}{\tau_{c}%
}\right)  t}\prod\limits_{i=1}^{N}e^{-\frac{1}{2}\frac{kT}{m}t^{2}q_{i}^{2}},\\
&  =\int_{0}^{\infty}dte^{-\left(  s+\frac{1}{\tau_{c}}\right)  t-\frac{1}%
{2}\frac{kT}{m}t^{2}q^{2}}.%
\end{align}
Finally performing the Laplace transform we find,
\begin{align}
F_{N}(\mathbf{q},s) &  =\sqrt{\frac{\pi m}{2kTq^{2}}}e^{\frac{m}{2kT}%
\frac{\left(  s+\frac{1}{\tau_{c}}\right)  ^{2}}{q^{2}}}\mathrm{erfc}\left(
\sqrt{\frac{m}{2kT}}\frac{s+\frac{1}{\tau_{c}}}{q}\right),  \nonumber\\
&  =\sqrt{\frac{\pi m}{2kTq^{2}}}e^{z^{2}}\mathrm{erfc}\left(  z\right)
\equiv F\left(  q,z\right),  \label{111}%
\end{align}

where,
\begin{equation}
z(q,s)=\sqrt{\frac{m}{2kT}}\frac{1}{\tau_{c}}\frac{\left(  1+s\tau_{c}\right)
}{q}. \label{AAA}%
\end{equation}

We have been working with the Laplace transform of various functions of time.
This implies that these functions are causal i.e. equal to zero for $t<0.$ If
we make the replacement $s\rightarrow i\omega$, and take two times the real
part of the resulting expression, the results will apply to the even $\left(
f\left(  -t\right)  =f\left(  t\right)  \right)  $ extension of the causal
functions in agreement with other authors e.g. \cite{mcgregor}. Unless
specified it should be assumed that a spectrum refers to the even extension.
From now on we use,
\begin{equation}
F\left(  q,\omega\right)  =F\left(  q,z\left(  q,s=i\omega\right)  \right).
\end{equation}

It is immediately seen that the result is independent of the number of
dimensions N. The dimensionality of the model only appears in $q$ where,
\[
q^{2}=\sum\limits_{i=1}^{N}q_{i}^{2}.%
\]
We note that $q$ can never be negative, this is important to remember when
integrating and/or summing over discrete values of $q.$ However, $q_{i},$ a
single component of $\mathbf{q}$ can be negative. The conditional density for
any number of dimensions (\ref{11}) can be written as,
\begin{equation}
p(q,\omega)=2\operatorname{Re}\left[  \frac{F(q,\omega)}{1-\frac{1}{\tau_{c}%
}F(q,\omega)}\right].  \label{eq:pdf}%
\end{equation}

Thus, we observe agreement for the spectrum of the conditional probability
given by the CTRWT for 1, 2, and 3 dimensions. Furthermore, there are no cross
correlations between the different directions in Cartesian coordinates,
therefore one can compute values of a higher dimensional model from a lower
dimensional model, given that this model was projected from Cartesian
coordinates. Assuming Cartesian coordinates and given no cross-correlation in
the components of the functions being correlated we can compute a 3D result
from three 1D results, or one 2D result and one 1D result. In the latter case
the 2D model can include functions with cross correlation.

\subsubsection{\bigskip Comparison with diffusion theory.
\label{sec:diffusecompare}}

To compare to diffusion theory we define a length scale and ballistic
collision time, naturally the ballistic time should scale linearly with the
length, and inversely with the thermal speed of the system, thus $\tau
_{b}=L\sqrt{\frac{m}{kT}}$. For diffusion theory to be valid we must have
$\tau_{b}/\tau_{c}\gg1$ and $1\gg\omega\tau_{c}$ so that $z$ becomes very
large for not too large $q$,
\[
z\approx\sqrt{\frac{m}{2kT}}\frac{1}{\tau_{c}}\frac{1}{q}=\frac{1}{\sqrt{2}%
}\frac{\tau_{b}}{\tau_{c}}\frac{1}{qL_{x}}>>1.
\]
For large z the asymptotic expansion for the complimentary error function can
be used,
\begin{equation}
\mathrm{erfc}(z)\rightarrow\frac{e^{-z^{2}}}{\sqrt{\pi}z}\left(  1-\frac
{1}{2z^{2}}\right). \label{eq:erfexp}%
\end{equation}
For now, we keep the full form of $z$ prior to expansion of the error
function, and substitute equation~(\ref{eq:erfexp}) into
equation~(\ref{eq:pdf}),
\begin{equation}
p(q,s=i\omega)=2\operatorname{Re}\left(  \frac{(1+i\omega\tau_{c})^{2}}%
{\frac{kT}{m}q^{2}\tau_{c}+i\omega(1+i\omega\tau_{c})^{2}}\right).  \label{bg4}%
\end{equation}

We then take the diffusion limit $\left(  \omega\tau_{c}<<1\right)  $ with the
result,
\begin{equation}
p\left(  q,\omega\right)  =2\operatorname{Re}\left(  \frac{1}{\frac{kT}%
{m}q^{2}\tau_{c}+i\omega}\right)  =2\operatorname{Re}\left(  \frac{1}{\left(
Dq^{2}\right)  +i\omega}\right). \label{BG2}%
\end{equation}

Since $D_{N}=\frac{\left\langle v_{N}^{2}\right\rangle }{N}\tau_{c}$ where N
is the number of dimensions and $\left\langle v_{N}^{2}\right\rangle
=N\frac{kT}{m},$ we have inserted the diffusion coefficient,
\begin{align}
D=\tau_{c}\frac{kT}{m}. \label{eq:tauctodiff}%
\end{align}
Equation~(\ref{BG2}) is immediately seen to be the Fourier transform of the
Green's function of the diffusion equation.

\subsection{Vector velocity autocorrelation function in an infinite domain.}

The vector velocity autocorrelation function can be written as an integration
over the vector components of velocity, analogous to the one dimensional
treatment in $\cite{zab}$,%

\begin{align}
S_{vv}\left(  \mathbf{q,}s\right)  =%
%TCIMACRO{\diint \limits_{-\infty}^{\infty}}%
%BeginExpansion
{\displaystyle\iint\limits_{-\infty}^{\infty}}
%EndExpansion
\mathbf{v\cdot v}_{0}\widehat{p}_{\mathbf{vv}_{0}}\left(  \mathbf{q},s\right)
\alpha(\mathbf{v}_{0})d^{3}\mathbf{v}_{0}d^{3}\mathbf{v}.%
\end{align}

Where $p_{\mathbf{vv}_{0}}\left(  \mathbf{q},s\right)  ~$is the $\cite{zab}$
Fourier-Laplace transform of the conditional probability for a particle which
has velocity $\mathbf{v}_{o}$ at $\left(  \mathbf{x}=0\mathbf{,}t=0\right)  $
to have the velocity $\mathbf{v}$ at $\left(  \mathbf{x,}t\right)  $ and satisfies,

\begin{equation}
p_{\mathbf{vv}_{0}}\left(  \mathbf{q},s\right)  =G_{v}\left(  \mathbf{q}%
,s\right)  \delta(\mathbf{v}-\mathbf{v}_{0})+\alpha(\mathbf{v)}G_{\mathbf{v}%
}\left(  \mathbf{q},s\right)  \rho_{o}(\mathbf{q,}s),%
\end{equation}

where,
\begin{equation}
\rho_{o}(\mathbf{q,}s\mathbf{)}=\frac{g(\mathbf{q},s,\mathbf{v}_{o}\mathbf{)}%
}{1-\int\alpha(\mathbf{v})g(\mathbf{q},s,\mathbf{v)d}^{3}\mathbf{v}},%
\end{equation}

is the Laplace-Fourier transform of the scattering density at $\left(
\mathbf{x,}t\right)  $ of particles that started at$\left(  \mathbf{x}%
=0\mathbf{,}t=0\right)  $ with velocity $\mathbf{v}_{o}.$ Then,
\begin{equation}
p_{\mathbf{vv}_{0}}\left(  \mathbf{q},s\right)  =G_{v}\left(  \mathbf{q}%
,s\right)  \delta(\mathbf{v}-\mathbf{v}_{0})+\frac{\alpha(\mathbf{v)}%
G_{\mathbf{v}}\left(  \mathbf{q},s\right)  g(\mathbf{q},s,\mathbf{v}%
_{o}\mathbf{)}}{1-\int\alpha(\mathbf{v})g(\mathbf{q},s,\mathbf{v)d}%
^{3}\mathbf{v}},%
\end{equation}
and using as above, $\frac{1}{\tau_{c}}G_{\mathbf{v}}\left(  \mathbf{q}%
,s\right)  =g_{\mathbf{v}}\left(  \mathbf{q},s\right)  $ we have,
\begin{align}
S_{\mathbf{vv}}=\int\mathbf{v}^{2}G_{v}\left(  \mathbf{q},s\right)
\alpha\left(  \mathbf{v}\right)  d^{3}\mathbf{v}+~\frac{\frac{1}{\tau_{c}%
}\left(  \int\mathbf{v}\alpha\left(  \mathbf{v}\right)  G_{v}(\mathbf{q,}%
s)d^{3}\mathbf{v}\right)  ^{2}}{1-\frac{1}{\tau_{c}}\int\alpha\left(
\mathbf{v}\right)  G_{v}(\mathbf{q,}s)d^{3}\mathbf{v}}.%
\end{align}
For simplicity we write this equation as
\begin{align}
S_{\mathbf{vv}}(\mathbf{q,}s)=H+\frac{\frac{1}{\tau_{c}}\mathbf{K}^{2}%
}{1-\frac{1}{\tau_{c}}L}.%
\end{align}

Where,%

\begin{align}
H  &  =\int v^{2}G_{v}\left(  \mathbf{q},s\right)  \alpha\left(  \mathbf{v}\right)
d^{3}\mathbf{v},\\
L  &  =\int G_{v}\left(  \mathbf{q},s\right)  \alpha\left( \mathbf{v}\right)
d^{3}\mathbf{v},\\
\mathbf{K}  &  =\int\mathbf{v}G_{v}\left(  \mathbf{q},s\right)  \alpha\left(
\mathbf{v}\right)  d^{3}\mathbf{v},\\
G(\mathbf{q,}s\mathbf{)}  &  =\frac{1}{\left(  s+\frac{1}{\tau_{c}}\right)
+i\mathbf{q\cdot v}},%
\end{align}

Carrying out all the integrations we find:%

\begin{equation}
S_{vv}(q,s)=\frac{z+\sqrt{\pi}e^{z^{2}}(1-z^{2})\mathrm{erfc}\left(  z\right)
}{\sqrt{\frac{m}{2kT}}q}-\frac{\lambda}{q^{2}}\frac{\left(  \sqrt{\pi
}ze^{z^{2}}\mathrm{erfc}\left(  z\right)  -1\right)  ^{2}}{1-\sqrt{\frac{\pi
m}{2kT}}\frac{1}{\tau_{c}q}e^{z^{2}}\mathrm{erfc}\left(  z\right)  }.
\label{eq:vectorvelcorr}%
\end{equation}

This is the Fourier Laplace transform for the velocity autocorrelation
function of an unbounded continuous time random walk in 3D, given a Maxwell
velocity distribution, thermalizing gas collisions and Poisson distributed
collision times. \ 

To calculate the spectrum of the position averaged velocity autocorrelation
function we take the limit as $q\rightarrow0,(z>>1)$ and we use the asymptotic
expansion of the$~\mathrm{erfc(z)},$ equation (\ref{eq:erfexp}), where we must take the
expansion to the second term, %

\begin{equation}
S_{vv}\left(  \omega\right)  =\lim_{q\rightarrow0}\left[
\begin{array}
[c]{c}%
\frac{z+(1-z^{2})\frac{1}{z}(1-\frac{1}{2z^{2}})}{\sqrt{\frac{m}{2kT}}q}\\
-\frac{1}{\tau_{c}q^{2}}\frac{\left(  \frac{1}{2z^{2}}\right)  ^{2}}%
{1-\sqrt{\frac{m}{2kT}}\frac{1}{\tau_{c}q}\frac{1}{z}(1-\frac{1}{2z^{2}})}%
\end{array}
\right]  .
\end{equation}

The second term in the sum does not contribute. This is expected from the
derivation in reference $\cite{zab}$, and signifies that the scattered
trajectories do not contribute to the VACF because of cancellation when
averaging over direction of the scattered particles. The first term can be
simplified, and the limit taken,
\begin{align}
S_{vv}\left(  \omega\right)   &  =\lim_{q\rightarrow0}\frac{\frac{3}{2z}%
-\frac{1}{2z^{3}}}{\sqrt{\frac{m}{2kT}}q},\\
&  =\frac{3kT}{m}\frac{1}{\left(  s+\frac{1}{\tau_{c}}\right)  }.%
\end{align}
Which has an inverse Laplace transform,
\begin{equation}
R_{vv}(t)=\frac{3kT}{m}e^{-\frac{t}{\tau_{c}}}. \label{AA}%
\end{equation}
This is the expected velocity autocorrelation function for an infinite domain.
This particular result could have been found by finding the single component
spectrum and multiplying by the number of dimensions, for this case the
information in $q$ contained in equation~(\ref{eq:vectorvelcorr}) is removed
by the average over position $\left(  q\rightarrow0\right)  $.

\section{Stochastic motion in bounded domains}

In section~\ref{ctrwt} we obtained a general expression for the propagator of our
CTRWT model in an infinite domain. In this sction we show how our result
(\ref{eq:pdf}) may be used to construct the spectrum of the position and
velocity auto-correlation functions in a bounded domain.

\subsection{Spectrum of the position and velocity auto-correlation functions
in a bounded domain}

We consider stochastic motion within a rectangular domain of the size
$\{$$L_{x}$, $L_{y},$ $L_{z}$$\}.$ \ Using our method of images $\cite{swank}$
each reflection from a boundary is replaced by a particle coming from an image
source, the original particle being considered as leaving the physical bounded
region. So the probability of arriving at a given point, P$\left(  x,t\right)
$ say, is given by the sum of probabilities of arriving from the original,
physical source and all the image sources. As time increases more distant
image sources come into play. Physically it is similar to standing between two
perfect facing mirrors.

For each source point there is a set of image points, one each in a lattice of
repetitions of the physical domain.

When we use this probability function to calculate averages of functions of
position the probability $\mathrm{P}\left(  x,t\right)  $ has to be averaged
over all possible starting positions in the physical cell and summed over all
image points. This is equivalent to integrating the infinite domain
probability function over all possible source points. For a more complete
description refer to~$\cite{swank,oppenheim1964,wayne1966,tarczon1985}.$

The procedure can be clearly seen, for example, by considering the position
coordinate as the function to be averaged. In this example as we go along the
coordinate (in the positive direction) in the physical cell the coordinate
increases. As we cross the boundary into the image cell the image coordinate
reaches a maximum at the boundary and then decreases (negative slope). If we
continue in this fashion the coordinate function will be a simple triangle
wave, zero at the origin (asymmetric), with amplitude $L_{x}/2$, and period
$2L_{x},$ the size of the physical cell being given by $L_{x}$.

Periodic functions can be represented in a Fourier series, for the triangle
wave representing the position coordinate in the image cells centered at the
origin we have,
\begin{equation}
~\widetilde{x}\left(  x\right)  =\sum_{n=\mathrm{odd}}-i^{n}\frac{2L_{x}}{\pi^{2}n^{2}}e^{i\frac{\pi n}{Lx}x}. \label{eq:repeatingx}%
\end{equation}

The spectrum of the position correlation function can thus be written in terms
of the periodic varying function, $\widetilde{x}$%,

\begin{align}
S_{xx}\left(  \omega\right)  =\frac{1}{8 \pi^{3} L_{x} L_{y} L_{z}}
\int_{-\mathbf{L}/2}^{\mathbf{L}/2} d^{3}\mathbf{x}_{0} \int_{-\infty}^{\infty
}d^{3}\mathbf{x}\int_{-\infty}^{\infty}d^{3}\mathbf{q}~\widetilde{x}%
x_{0}p(q,\omega)e^{-i\mathbf{q}\cdot(\mathbf{x}-\mathbf{x}_{0})}%
,\label{eq:3Dstart}%
\end{align}

where$~p(q,\omega)$ is the conditional density found in equation~(\ref{eq:pdf}%
). Integration over $y$ and $z$ gives $2\pi\delta(q_{y})$ and $2\pi
\delta(q_{z})$, respectively. Due to the normalization in equation
\ref{eq:3Dstart}, subsequent integration over $y_{0}$ and $z_{0}$ will give
unity. Integration over $x$ gives $\sum_{n}-i^{n}\frac{2L_{x}}{\pi^{2}n^{2}%
}2\pi\delta\left(  q_x-\frac{\pi n}{L_{x}}\right)  ,$ due to the definition of
$\widetilde{x}~$in equation~(\ref{eq:repeatingx}) where the sum is over $n$,
where $n$ are odd integers,%

\begin{equation}
S_{xx}\left(  \omega\right)  =\frac{1}{L_{x}}\sum_{n=\mathrm{odd}}-i^{n}\frac{2L_{x}}{\pi^{2}n^{2}}\int_{-L/2}^{L/2}%
dx_{0}~x_{0}p\left(q=|q_x|=\frac{\pi |n|}{L_{x}},\omega\right)e^{i\frac{\pi n}{L_{x}}x_{0}}.%
\end{equation}

Integration over $x_{0},$ and taking the even extension of the causal function
as above, yields,
\begin{equation}
S_{xx}\left(  \omega\right)  =\sum_{n=\mathrm{odd}}\frac{4L_{x}^{2}}{\pi^{4}n^{4}}  p\left(|q_x|=\frac{\pi |n|}{L_{x}},\omega\right)
.\label{eq:BG3}%
\end{equation}
For the velocity autocorrelation function we have,%
\begin{equation}
S_{vv}\left(  \omega\right)  =\omega^{2}\sum_{n=\mathrm{odd}}\frac{4L_{x}^{2}}{\pi^{4}n^{4}} p\left(\frac{\pi |n|}{L_{x}%
},\omega\right)  .\label{BG4}%
\end{equation}

\subsection{Long time tails arise in diffusive motion in bounded
domains.\label{sec:longtimetail}}

The series expansion (\ref{eq:BG3}) for the spectrum of the position
auto-correlation function is universal, it is valid for any mean free path
from the quasi-ballistic ($\xi=\tau_{c}/\tau_{b}>>$1) to the diffusive
($\xi<<$1) regime of motion. In this section we obtain closed-form expressions
valid in the diffusive regime.

We start from equation~(\ref{eq:pdf}),%

\begin{equation}
p\left(  q,\omega\right)  =2\mathrm{Re}\left[\frac{\sqrt{\frac{m\pi}{2kT}}\frac{1}{q}%
\text{e}^{z^{2}}\text{erfc}\left(  z\right)  }{1-\sqrt{\frac{m\pi}{2kT}}%
\frac{1}{q\tau_{c}}\text{e}^{z^{2}}\text{erfc}\left(  z\right)  }\right]=2\mathrm{Re}\left[\tau
_{c}\left(  \frac{1}{1-\sqrt{\frac{m\pi}{2kT}}\frac{1}{q\tau_{c}}%
\text{e}^{z^{2}}\text{erfc}\left(  z\right)  }-1\right) \right], \label{12}%
\end{equation}

where $z$ is given by (\ref{AAA}).

Assuming that the propagator is the even extension of the casual conditional
probability $P(x,t),$ the spectrum of the position auto-correlation function
in a finite system of size $L_{x}\text{ reads }$ \cite{swank}, (\ref{eq:BG3})%

\begin{equation}
S_{xx}\left(  \omega\right)  =\frac{8L_{x}{}^{2}}{\pi^{4}}\sum
_{n=1,3\text{...}}^{\infty}\frac{  p\left(  q_{n},\omega\right)
  }{n^{4}},\label{13}%
\end{equation}

where,
\begin{align}
z_{n}  &  =\sqrt{\frac{m}{2kT}}\frac{1+i\omega\tau_{c}}{q_{n}\tau_{c}},\\
q_{n}  &  =\frac{n\pi}{L_{x}}.%
\end{align}

As shown in appendix~\ref{longtimetailderiv} the result is,

\begin{equation}
S_{xx}\left(  \omega^{\prime}\right)  =\frac{2\tau_{c}L_{x}{}^{2}\xi^{2}%
}{\omega^{\prime}{}^{2}\left(  1+\omega^{\prime}{}^{2}\right)  }%
(1-\Delta\lbrack\xi,\omega^{\prime}])
\end{equation}
where $\xi=\tau_{c}/\tau_{b}$, $\omega^{\prime}=\omega\tau_{c}$, and%

\begin{equation}
\Delta\left[  \xi,\omega^{\prime}\right]  =\frac{\sqrt{2}\xi}{\left(
\omega^{\prime}\right)  ^{1/2}\left(  1+\omega^{\prime}{}^{2}\right)  }%
\frac{\sin\left[  \frac{\sqrt{\omega^{\prime}}\left(  1+\omega^{\prime
}\right)  }{\sqrt{2}\xi}\right]  \left(  1+2\omega^{\prime}-\omega^{\prime}%
{}^{2}\right)  +\sinh\left[  \frac{\left(  1-\omega^{\prime}\right)
\sqrt{\omega^{\prime}}}{\sqrt{2}\xi}\right]  \left(  1-2\omega^{\prime}%
-\omega^{\prime}{}^{2}\right)  }{\cos\left[  \frac{\sqrt{\omega^{\prime}%
}\left(  1+\omega^{\prime}\right)  }{\sqrt{2}\xi}\right]  +\cosh\left[
\frac{\left(  -1+\omega^{\prime}\right)  \sqrt{\omega^{\prime}}}{\sqrt{2}\xi
}\right]  }. \label{as7}%
\end{equation}

Going back to original variables $\omega$,$\tau_{c}$,$\tau_{b}$ in the
prefactor, noting that $\left(  \frac{L_{x}}{\tau_{b}}\right)  ^{2}=kT/m$ and
using $D$=$\frac{kT}{m}\tau_{c}$, ($D$ is the diffusion coefficient), we find,

\begin{equation}
S_{xx}\left(  \omega\right)  =\frac{2D}{\omega^{2}\left(  1+\left(  \omega
\tau_{c}\right)  {}^{2}\right)  }\left(  1-\Delta\left[  \xi,\omega\tau
_{c}\right]  \right).  \label{as8}%
\end{equation}

Note, that our result (\ref{as7},\ref{as8}) for the spectrum of the position
correlation function is valid for any $\omega$ from 0 to $\infty$ as long as
$\tau_{c}$$<<$$\tau_{b}$, which is the condition for diffusive motion. Taking
the limit $\omega$ $\rightarrow$ 0 we obtain,%

\begin{equation}
S_{xx}\left(  0\right)  =\frac{L_{x}^{4}}{60D}-\frac{1}{6}L_{x}^{2}\tau
_{c}=\frac{L_{x}^{4}}{60D}\left(  1-10\xi^{2}\right).
\end{equation}
which is valid for the ``non-adiabatic''\ regime of motion $\omega$ $<<$
1/$\tau_{d}$. Here, the first term is well known from ``classical''\ diffusion
theory, \cite{mcgregor}. The second term, $\sim$ $10\xi^{2}$ is the next order
correction from our CTRWT model. When diffusion theory is valid the correction
is very small. \newline We see that the prefactor in (\ref{as8}) does not
involve any information on the system size, while the term $\Delta$($\xi
$,$\omega$') \ depends on the size of the system. Thus the prefactor
represents the spectrum for an infinite system $S_{xx}^{\infty}$ while the
term $\Delta$($\xi$,$\omega$') is a correction due to the finite size,%

\begin{equation}
S_{\text{xx}}(\omega) =S_{\text{xx}}^{\infty}(\omega)\left(  1-\Delta\left[
\xi,\omega\tau_{c}\right]  \right),
\end{equation}%

\begin{equation}
S_{\text{xx}}^{\infty}(\omega) =\frac{2D}{\omega^{2}\left(  1+\omega^{2}%
\tau_{c}{}^{2}\right)  }. \label{asInf}%
\end{equation}

\begin{figure}[ptb]
\begin{center}
\includegraphics[width=0.7\textwidth]{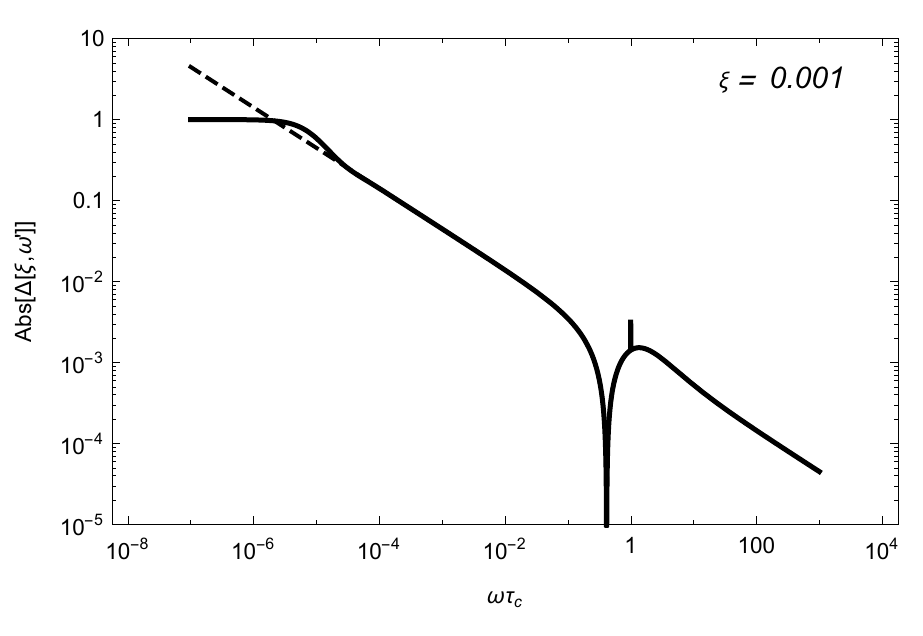}
\end{center}
\caption{Correction to the spectrum of position correlation function
(\ref{as7}) due to the finite size of the system, (solid line). Approximate
correction (\ref{as9}), (dashed line). For $\omega\tau_{c}<<1$the correction
is negative, it has irregularities in the vicinity of $\omega\tau_{c}\sim1$.
However, the detailed behavior of the correction for $\omega\tau_{c}\gtrsim1 $
is of minor importance since its relative magnitude is small.}%
\label{fig:AP1}%
\end{figure}

We may \ greatly simplify (\ref{as7}) noting that for large values of the
argument $\frac{\sqrt{\omega\tau_{c}}(1-\omega\tau_{c})}{\sqrt{2}\xi}$$>>1$
the hyperbolic functions dominate and (\ref{as7}) reduces to:%

\begin{equation}
\Delta(\xi,\omega\tau_{c})\approx\frac{\sqrt{2}\xi\left(  1-2\omega\tau
_{c}-\omega^{2}\tau_{c}{}^{2}\right)  }{\left(  \omega\tau_{c}\right)
^{1/2}\left(  1+\omega^{2}\tau_{c}{}^{2}\right)  }. \label{as9}%
\end{equation}
We illustrate the frequency dependence of the finite size corrections
(\ref{as7}), solid line, and (\ref{as9}), dashed line, on Figure~\ref{fig:AP1}.
 
For $\omega\tau_{c}<<1$, the condition $\frac{\sqrt{\omega\tau_{c}}%
	(1-\omega\tau_{c})}{\sqrt{2}\xi}$$>>$1 reduces to $\omega\tau_{c}>>2\xi^{2}$,
or $\omega\tau_{d}>>2\pi^{2}$. This condition together with $\omega\tau_{c}%
$$<<$1 constitutes the \textquotedblleft classical\textquotedblright
\ conditions for the \textquotedblleft adiabatic\textquotedblright\ regime of
spin-motion: {\ }$1/\tau_{d}<< \omega<< 1/\tau_{c}$. When these conditions are fulfilled,

\begin{equation}
\Delta(\xi,\omega\tau_{c})\approx\frac{\sqrt{2}\xi}{\left(  \omega\tau
	_{c}\right)  ^{1/2}}.%
\end{equation}

\begin{figure}[ptb]
\begin{center}
\includegraphics[width=0.7\textwidth]{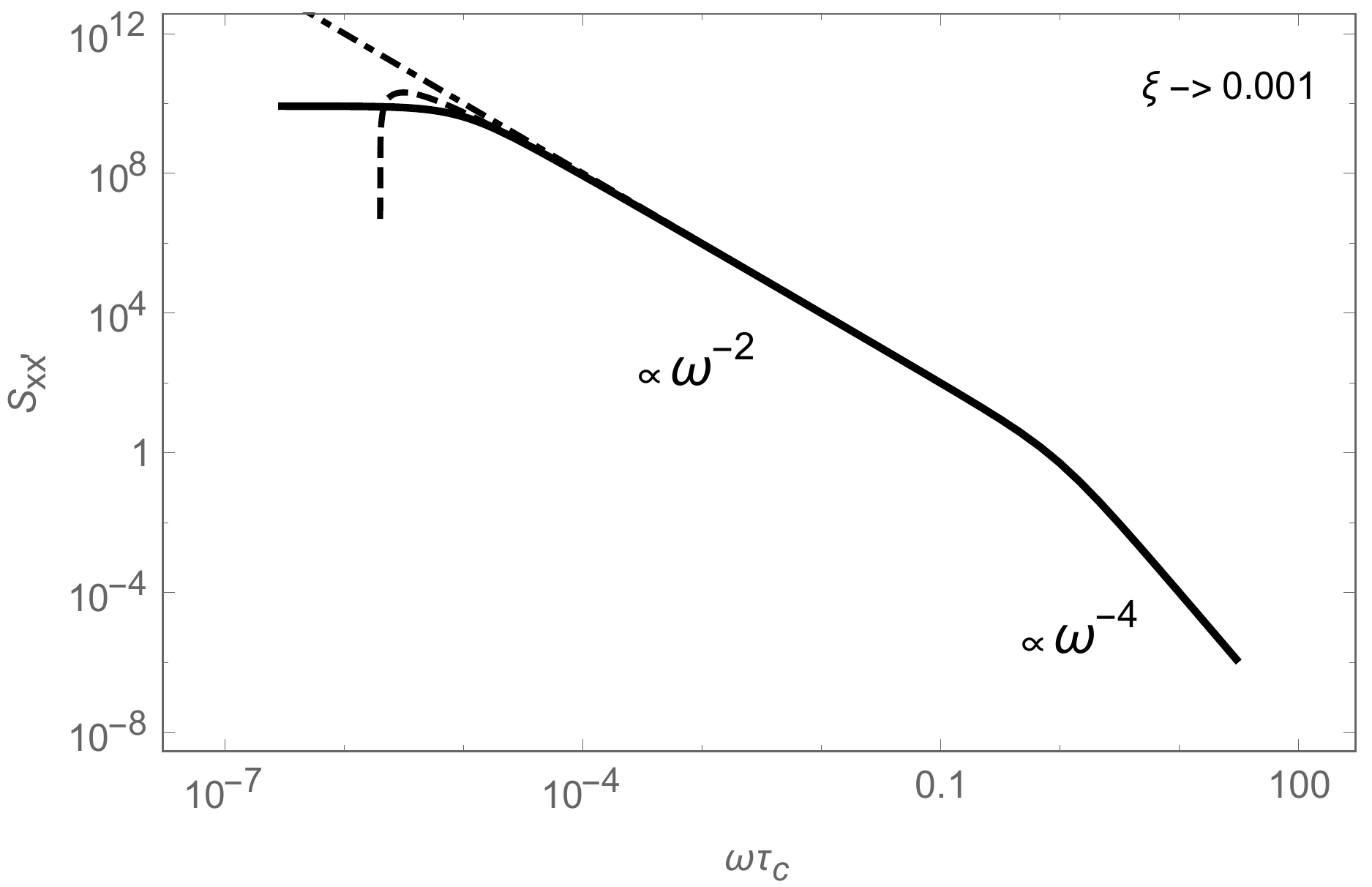}
\end{center}
\caption{Normalized position correlation spectrum $S_{\text{xx}}%
(\omega)/(2D\tau_{c}{}^{2})$ calculated from (\ref{as8}), using the exact
finite size correction (\ref{as7}) (solid line), using the approximate
correction (\ref{as9}), (dashed line), and for the infinite system
(\ref{asInf}), (Dot-Dashed line). Saturation of the spectrum for $\omega
\tau_{d}<1$ is a signature of the finite size.}%
\label{fig:AP2}%
\end{figure}

From analysis of Figure~\ref{fig:AP1} we conclude that the
expression~(\ref{as9}) indeed gives an excellent approximation to the exact
result (\ref{as7}) for the ``adiabatic{''} and {``}super-adiabatic{''} regimes
of spin-motion. Figure~\ref{fig:AP2} shows the normalized position correlation
\ spectrum \ $S_{xx^{\prime}}$($\omega$)/($D\tau_{c}{}^{2}$) calculated
from~(\ref{as7}, \ref{as8}) \ as well as using approximation~(\ref{as9}).

\subsubsection{ Spectrum of the correlation function of a single velocity
component and long time memory in finite systems.}

In this section we show the existence of a long time tail in the velocity
autocorrelation function that arises only in the bounded domain. In free space
it was shown in equation~(\ref{AA}) that there is no long time tail, and the velocity
autocorrelation decays exponentially in time.

The spectrum of the correlation function of a single velocity component
$S_{v_{x}v_{x}}(\omega)$ may be found from,%

\begin{equation}
S_{v_{x}v_{x}}(\omega)=\omega^{2}S_{xx}(\omega),
\end{equation}

which gives,%

\begin{equation}
S_{v_{x}v_{x}}(\omega)=S_{v_{x}v_{x}}^{\infty}(\omega)(1-\Delta( \xi
,\omega\tau_{c})), \label{as10}%
\end{equation}%

\begin{equation}
S_{v_{x}v_{x}}^{\infty}(\omega)\approx\frac{2D}{\left(  1+\omega^{2}\tau_{c}%
{}^{2}\right)  }, \label{as11}%
\end{equation}

where the finite size correction, $\Delta(\xi,\omega\tau_{c}),$ is the same as
for the spectrum of the position correlation function.

\begin{figure}[ptb]
\begin{center}
\includegraphics[width=0.7\textwidth]{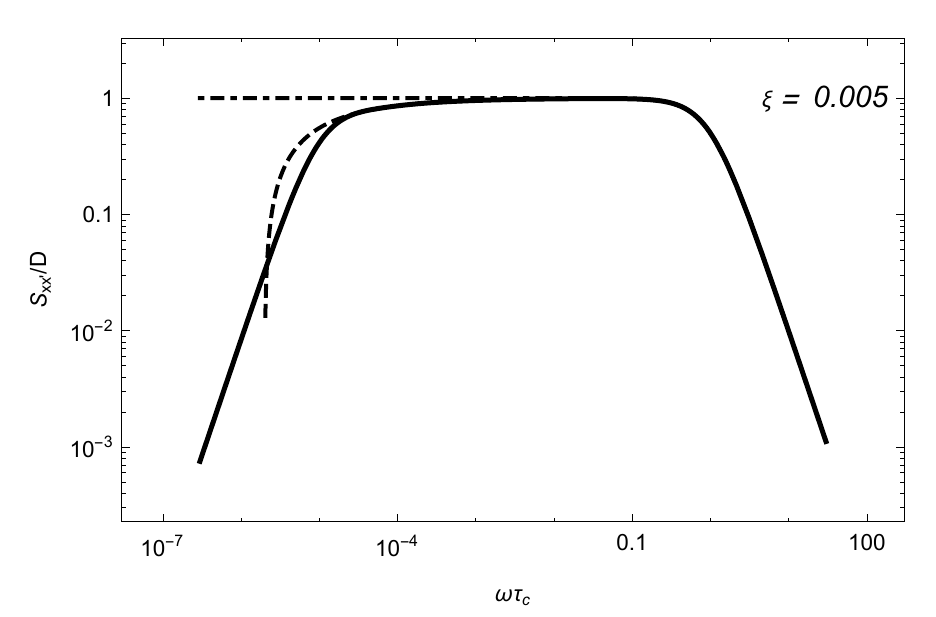}
\end{center}
\caption{Normalized spectrum of the correlation function for a single velocity
component $S_{v_{x}v_{x}}/D$ calculated according to (\ref{as10}), using the
exact finite size correction (\ref{as7}) (solid line), using the approximate
correction (\ref{as9}) (dashed line), and for the infinite system
(\ref{as11}), (dot-dashed line). The low frequency cut-off for $\omega
<<\xi^{2}$ is a signature of finite size systems.}%
\label{fig:AP3}%
\end{figure}

Figure~\ref{fig:AP3} shows the exact spectrum of the correlation function of
a single velocity component given by (\ref{as7}, \ref{as10}, \ref{as11}), solid
line, as well as the spectrum obtained using the approximation (\ref{as9}),
dashed line, and the spectrum for an infinite domain (\ref{as11}), dot-dashed
line. 

Applying the inverse Fourier transform to (\ref{as11}) \ we recover the well
known result for the velocity correlation function for an infinite domain
(\ref{AA}):%

\begin{equation}
R_{v_{x}v_{x}}^{\infty}(t)=\frac{kT}{m}e^{-\left|  t\right|  \left/  \tau
_{c}\right.  }.\label{as13}%
\end{equation}

The inverse Fourier transform for the exact result (\ref{as7}, \ref{as10},
\ref{as11}) is unknown. However, the relatively simple form of (\ref{as9})
allows an inverse Fourier transformation. For $\tau_{d}$ $>>$ t $>>$$\tau_{c}$
we obtain,

\begin{equation}
R_{v_{x}v_{x}}(t)\approx R_{v_{x}v_{x}}^{\infty}(t)-\frac{2}{\sqrt{\pi}}%
\xi\frac{kT}{m}\left(  \frac{t}{\tau_{c}}\right)  {}^{-1/2}. \label{as12}%
\end{equation}

\begin{figure}[ptb]
\begin{center}
\includegraphics[width=0.7\textwidth]{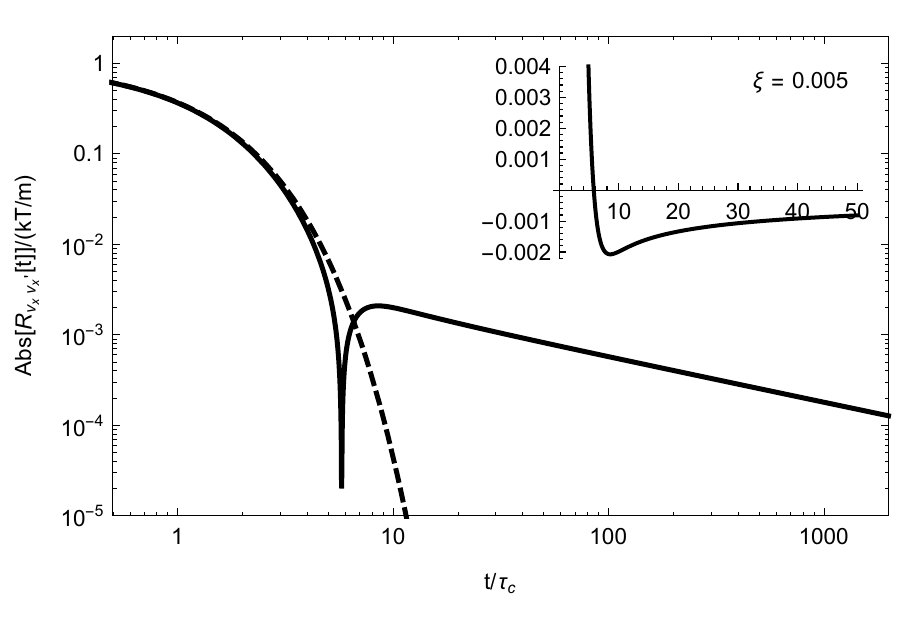}
\end{center}
\caption{Normalized velocity correlation function. From the analytic result
(\ref{as12}) valid for $\tau_{c}<<t<<\tau_{d}$, solid line. For the infinite
system (\ref{as13}), dashed line. Negative spikes correspond to positions
where functions charge their sign.}%
\label{fig:AP4}%
\end{figure}

One can see that the finite size correction has the form of a long-time tail
$\sim$$\left(  t\left/  \tau_{c}\right.  \right)  {}^{-1/2}$ with a relative
magnitude of the order of $\xi$ = $\tau_{c}$/$\tau_{b}$.\newline
\hspace*{0.5ex} Both solutions (\ref{as13}, \ref{as12}) are illustrated in
figure~\ref{fig:AP4}. 

For very short times, $t$ $<<$ $\tau_{c}$, all the results decay exponentially
with a time constant $\tau_{c}$. For longer times only $R_{v_{x}v_{x}}%
^{\infty}(t)$ decays exponentially. The velocity correlation function
$R_{v_{x}v_{x}}(t)$ for a finite system has a very different behavior: for
longer times it crosses the $\tau/\tau_{c}$ axis from positive values to
negative ones ( at the position where a spike is observed on the log-log
plot), and for even later times it goes back toward the $\tau/\tau_{c}$ axis
being negative. It stays negative for very long times, until $\tau$
approaches $\tau_{d}$, slowly decaying in magnitude according to a power law
$\sim t^{-1/2}$. It may be shown, see appendix (\ref{longtimetailderiv}), that for even longer times, $\tau\gg\tau_{d}%
$, it again decays exponentially with time constant $\tau_{d}$. This negative long time
tail, $\sim t^{-1/2},$ in the velocity correlation function is the same for 1 D, 2 D,
and 3 D systems and arises due to reflections from the system boundaries. This
is in agreement with the 1 dimensional treatment examined in reference
\cite{oppenheim1964}. This is in contrast to molecular dynamics where long
range hydro-dynamic forces continuously act on the trajectories. These forces
tend to produce cross correlation between motion in different directions, for
example vorticular motion. The cross-correlation is examined in
reference~\cite{keyes75}. See also \cite{Dib2006}. Including cross-correlation
of such kind leads to ''long time tails''\ \ $\propto$ $t^{-d/2}$, where $d$
represents the number of dimensions in the system. However, in systems of a
relative low density gas, the cross correlations resulting from hydro-dynamic
forces are suppressed, and our result completely describes the dynamics.

We can see the reason for the difference between our result and previous
results showing a tail $\left(  \sim t^{-d/2}\right)  $ by examining reference
\cite{keyes75}. Equation (1) in that paper, while calculating the correlation
function of a single velocity component, is a function of $k^{\prime2}%
=\sum_{i=1}^{d}k_{i}^{\prime2}.$ However in calculating the position
correlation function $\left\langle x_{i}\left(  0\right)  x_{i}\left(
\tau\right)  \right\rangle $ for each $i,$ the dependence on the other
components, $j\neq i$, integrates out due to normalization of the conditional
probability so the correlation $\left\langle x_{i}\left(  0\right)
x_{i}\left(  \tau\right)  \right\rangle $ only depends on $k_{i}$. Using
(\ref{1}) to get the correlation function of the individual velocity
components $v_{i}$ we see that these each depend only on its particular
$k_{i}$. The total velocity autocorrelation function $\left\langle
\mathbf{v}\left(  t\right)  \cdot\mathbf{v}\left(  t-\tau\right)
\right\rangle $ is then the sum of $d$ such terms. The result is then a sum of
terms $\left(  \sim t^{-1/2}\right)  .$ Equation (1) in \cite{keyes75}
contains products of functions of the different $k_{i}^{\prime}s$ and this
results in the $\left(  \sim t^{-d/2}\right)  $ behavior as shown in equations
(4), (16) and (18) of that work and implies correlations between the different
directions of motion which do not occur in diffusion theory.

Another aspect of the long-time tail is a non-linear behavior of the mean
square displacement,%

\begin{align}
<x^{2}(t)>=2\int_{0}^{t}(t-\tau)<v_{x}(0)v_{x}(\tau)>d\tau=  &  2\frac{kT}%
{m}\int_{0}^{t}\left(  e^{-t\left/  \tau_{c}\right.  }-\frac{2}{\sqrt{\pi}}%
\xi\left(  \frac{t}{\tau_{c}}\right)  {}^{-1/2}\right)  (t-\tau)d\tau,\\
&  \approx2Dt\left(  1-\frac{8}{3\sqrt{\pi}}\sqrt{\frac{t}{\tau_{d}}}\right).
\label{27}%
\end{align}

Recall that our results (\ref{as12}) and, hence, (\ref{27}) are valid for
$\tau_{c}$$<<$ $t$ $<<$ $\tau_{d}$. Again, we see that for not too long times
$t\gg\tau_{c}$ the diffusion is a stationary Markovian process with,%

\begin{equation}
<x^{2}(t)>=2Dt,\label{28}%
\end{equation}

while for longer times, but still { }$t$ $<<$ $\tau_{d}$, reflections from the
boundaries will alter the linear dependence (\ref{28}). While our correction
is small it is an indication of how the diffusion process eventually ends for
$t\gg\tau_{d}$ in a homogeneous steady state with,
\begin{equation}
\left\langle x^{2}\left(  t>>\tau_{d}\right)  \right\rangle =\left\langle
x^{2}\right\rangle.
\end{equation}

\textit{ }As our results for the propagator have been shown to agree with
diffusion theory in the limit $\xi=\frac{\tau_{c}}{\tau_{b}}\ll1,$ we expect
that the long time tail will also occur in the classical diffusion theory. We
show this in an appendix.

\subsection{Spectrum of the bounded domain position correlation function in
the ballistic or diffusive limits.}

\subsubsection{Diffusive Region}
In the diffusion limit, the spectrum of the position autocorrelation function is observed to
stratify into three regimes shown in figure \ref{fig:AP2}.

\begin{description}
	\item[a] {the non adiabatic regime, defined by $\omega<<\tau_{d}^{-1},$ with
		spectrum,}
	\begin{align}
	S_{xx}\left(  \omega\right)  \approx S_{xx}\left(  0\right)  \approx\frac{1}{
		60}\frac{L_{x}^{4}}{D}.
	\end{align}
	
	\item[b] {the adiabatic regime, defined by $\tau_{d}^{-1}<<$ $\omega
		<<\tau_{c}^{-1}, $ with spectrum,}
	\begin{align}
	S_{xx}\left(  \omega\right)  \approx\frac{2D}{\omega^{2}}.
	\end{align}
	
	\item[c] {the super-adiabatic regime, defined by $\tau_{c}^{-1}<<\omega,$
		with spectrum,}
	\begin{align}
	S_{xx}\left(  \omega\right)  \approx\frac{2D}{\omega^{4}\tau_c^2}.
	\end{align}
\end{description}

Where the diffusion coefficient, $D=\frac{k_{B}T}{m}\tau_{c},$ and the
diffusion time, $\tau_{d}=\frac{L^{2}\newline }{\pi^{2}D},$ the time constant
for the lowest diffusion mode.

\subsubsection{Quasi-Ballistic motion}

In this section we will discuss some general properties of the spectrum of the
position-position auto-correlation function (\ref{eq:BG3}) in the case of
quasi-ballistic motion.

With decreasing pressure, the diffusion time decreases and the collision time,
$\tau_{c},~$increases. The motion is no longer diffusive when the time to
cross the restricted volume approaches the collision time, $\tau_{b}=$
$\frac{L}{\sqrt{\frac{k_{b}T}{m}}}\lesssim\tau_{c}.$ This is considered to be
the quasi-ballistic region, and in this region there exists no adiabatic
regime, where $S_{xx}(\omega)\propto$ $\omega^{-2},$ only two distinct regions
are found.

 Considering the limit $\xi=\tau_{c}/\tau_{b}>>1$
with the spectrum given by equation (\ref{eq:BG3}) leads us to distinguish
three different regimes:\newline 

\begin{description}
\item[a] The non adiabatic regime or low frequency region, $\omega$$<<$
1/$\tau_{c}$$<<$1/$\tau_{b}$ and $|z_{n}|$ $<<$1. In this regime we can
replace $\text{exp}\left( z_{n}{}^{2}\right)  \text{erfc}\left( z_{n}\right)
$ in equations { (\ref{111}, \ref{eq:pdf})} by equation (\ref{eq:erfexp}), %
\[
p\left(  q_{n},\omega\right)  \approx 2 \left(  \frac{1}{1-\frac{1}{\sqrt{2\pi
}n\xi}}-1\right)  \approx\frac{2\tau_{c}}{n\sqrt{2\pi}\xi}\text{,}%
\]

with the spectrum of the position auto-correlation function given by equation
(\ref{eq:BG3}),
\begin{align}
S_{xx}\left(  \omega\right)  \approx\frac{16L_{x}{}^{2}}{\pi^{4}}\frac{1}%
{\sqrt{2\pi}\xi}\tau_{c}\sum_{n=1,3\text{...}}^{\infty}\frac{1}{n^{5}}%
=\frac{31L_{x}^{2}\tau_{b}\zeta\left(  5\right)  }{2\sqrt{2}\pi^{\frac{9}{2}}%
},
\end{align}
where $\zeta(n)$ is the Riemann zeta function.

\item[b] The intermediate regime defined by 1/$\tau_{c}$$<<$ $\omega$$<<$
$\sqrt{2}$$\pi$/$\tau_{b}$. In this regime $z_{n}$ is mostly imaginary but
still $|z_{n}|<<1$ and the low z expantion is valid leading to the same
spectrum as above.\newline 

\item[c] {the super-adiabatic regime or high frequency region, defined by
$\sqrt{2}\pi/\tau_{b}<<\omega,$ with spectrum found according to the large
$|z|>>1$ expansion, equation (\ref{eq:erfexp}), of the conditional density,
equation (\ref{eq:pdf}),}%

\begin{align*}
S_{\text{xx}}(\omega)  &  \approx\frac{2D}{\omega^{2}\left(  1+\left(
\omega\tau_{c}\right)  {}^{2}\right)  }\approx\frac{2D}{\omega^{4}\tau_{c}%
{}^{2}}\text{ ,}\\
S_{v_{x}v_{x}}(\omega)  &  \approx\frac{2D}{\left(  \omega\tau_{c}\right)
{}^{2}}.%
\end{align*}
\end{description}

\begin{figure}[ptb]
	\begin{center}
		\includegraphics[width=0.85\textwidth]{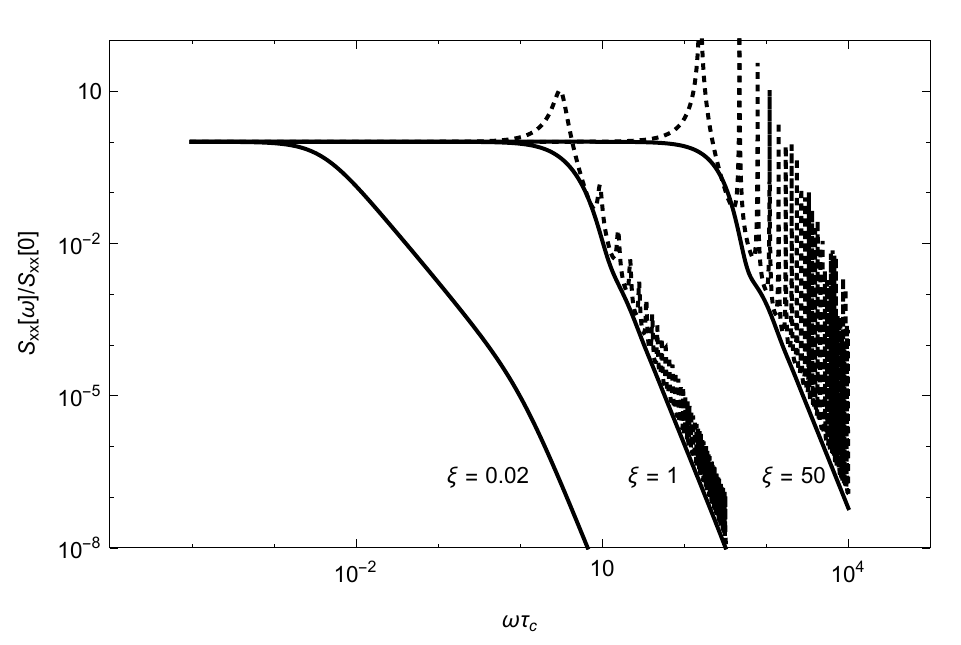}
	\end{center}
	\caption{The normalized spectrum of the position autocorrelation function, a
		comparison of the frozen picture with the thermalization picture in different
		regimes, diffusive ($\xi=0.02$), intermediate ($\xi= 1$), and quasi-ballistic
		($\xi=50$). Solid lines represent predictions of our CTRWT model, dotted lines
		correspond to the CTRW in the "frozen environment". For diffusive motion
		($\xi=0.02$) all models: CTRWT, CTRW in the "frozen environment", and
		classical diffusion theory give predictions which are indistinguishable in
		this plot. }%
	\label{fig:AP5}%
\end{figure}

Figure \ref{fig:AP5} shows spectra of the position correlation function for
different regimes of motion from diffusive to quasi - ballistic, it is
observed that scaling $\xi=\tau_{c}/\tau_{b}$ shifts the transition frequency
for the quasi-ballistic non-adiabatic to super-adiabatic regimes, when $\xi<1$
the diffusion region is observed. It is
interesting to compare the result for the spectrum of the position correlation
function given by our CTRWT model, equations~(\ref{eq:pdf},\ref{eq:BG3}), with
the prediction \cite{swank} for the CTRW in a {``}frozen environtment{''}.

Fig. \ref{fig:AP5} shows the evolution of the spectrum of position
auto-correlation function from the diffusive $\left(  \xi=.02\right)  $ to the
quasi-ballistic $\xi=50)$ regime of motion. Solid lines represent the
predictions of our thermalizing CTRWT model, dotted lines corresponds to the
model of CTRW in the { }{``}frozen invirontment{''} \cite{swank}.

For quasi - ballistic motion in the {``}frozen environtment{''} model
collisions with the boundaries lead to the formation of resonances, (for
details see \cite{swank}). In this model the character of the structure of the
resonances, as well as their width, depends on the parameter $\xi$ and the
number of dimensions in the the system. The higher is $\xi$ the more narrow
are the resonances. For the diffusive regime of motion ($\xi$ $<<$1) the
resonance structure is fully washed out and the prediction of all three models: our model of
CTRWT, CTRW in {``}frozen invirontment{''} and classical diffusion theory
\cite{mcgregor} are indistinguishable.

\begin{figure}[ptb]
\begin{center}
\includegraphics[width=0.7\textwidth]{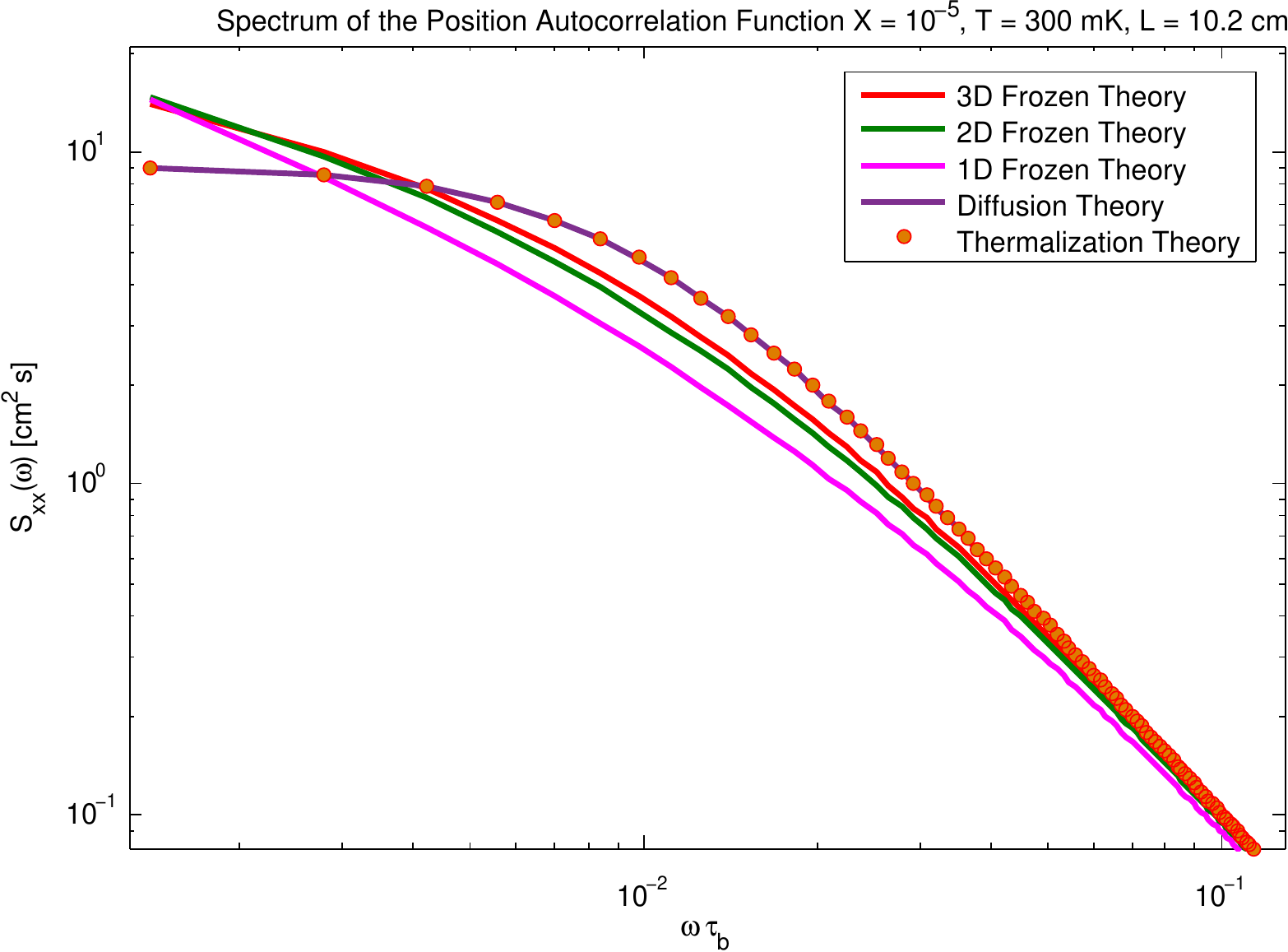}
\end{center}
\caption{The thermalization model compared to the ''frozen'' CTRW models, and
diffusion theory, in the diffusion limit, $\omega\tau_{c}<<1$. Diffusion
theory is accurate in this regime. We use the diffusion coefficient given by
reference \cite{opfer}. It is seen that the ''frozen'' CTRW models diverge
from diffusion theory at low frequencies. }%
\label{fig:Disagree}%
\end{figure}

$~$The resonances in the ballistic region would be smoothed out by averaging
the 'frozen scatterer' spectrum over a Maxwell-Boltzmann velocity
distribution, however in the zero frequency limit the velocity average of the
spectrum diverges. Furthermore velocity averaging the single velocity spectrum
in the diffusion region gives results which depend on the number of dimensions
as shown in figure~\ref{fig:Disagree}, in disagreement with the thermalization
model presented here.

\subsection{Application of the bounded domain correlation functions
\label{sec:bounddomain}}

Another correlation function of particular physical interest is the
position-velocity correlation function as it determines the frequency shift
linear in the electric field of spins precessing in magnetic and electric
fields. This is important in the search for electric dipole moments (edm),
where the presence of an edm results in frequency shifts which are also linear
in the applied electric field \cite{Lam05, pendlebury04}. With the use of
integration by parts, the frequency shift can be written it terms of the
imaginary component of the Fourier transform of the position auto-correlation
function, $\cite{pignol,swankThesis,pignol2015}$
\begin{align}
\delta\omega= &  -\omega\frac{\gamma^{2}E}{c}\mathrm{Im}\left[  \int
_{0}^{\infty}e^{-i\omega\tau}\left\langle B_{x}(t)x(t+\tau)+B_{y}%
(t)y(t+\tau)\right\rangle d\tau\right]  \nonumber\\
&  -\gamma^{2}\frac{E}{c}\left\langle B_{x}x+B_{y}y\right\rangle,
\label{eq:freqshift}%
\end{align}
here $E$ is the strength of the electric field appied in the $z$ direction,
and $B_{x,y}$ represents a perturbing magnetic field. $\left\langle
..\right\rangle $ represents an ensemble average. The frequency $\omega$ is
determined by the applied holding field $B_{0}$, also in the $z$ direction,
\begin{align}
\omega=\gamma B_{0}.
\end{align}
The field $B_{x}$ in equation~(\ref{eq:freqshift}) is a perturbation on the
holding field $B_{0}$ manifest from the inevitable inhomogeneities of
laboratory magnets. For accurate predictions of the relaxation and frequency
shifts accounting for linear and quadratic terms are enough $\cite{quartic}$,
any higher order terms are negligible. Due to the correlation between field
and position only asymmetric terms contribute, therefore only contributions
from linear inhomogeneities are required for an accurate prediction.
Therefore, we take,
\[
B_{x,y}\propto x,y,
\]
and the phase shift due to the $x~$component is proportional to the spectrum
of the position autocorrelation function,%
\begin{equation}
\delta\omega\propto\omega\mathrm{Im}\left[  S_{xx}\left(  \omega\right)
\right]  +\langle xx\rangle.
\end{equation}
In this case $S_{xx}(\omega)$ is the  spectrum obtained by using $p\left(
q,s=i\omega\right)$, where $p\left(
q,s\right)$ is the causal conditional density. A similar expression exists for the $y~$component.

The thermalization model of the random walk presented in this work is now used
to predict the phase shift of $^{3}$He, Larmor precessing in a dilute solution
in superfluid $^{4}$He, \cite{golub}.

\begin{figure}[ptb]
\begin{center}
\includegraphics[width=0.85\textwidth]{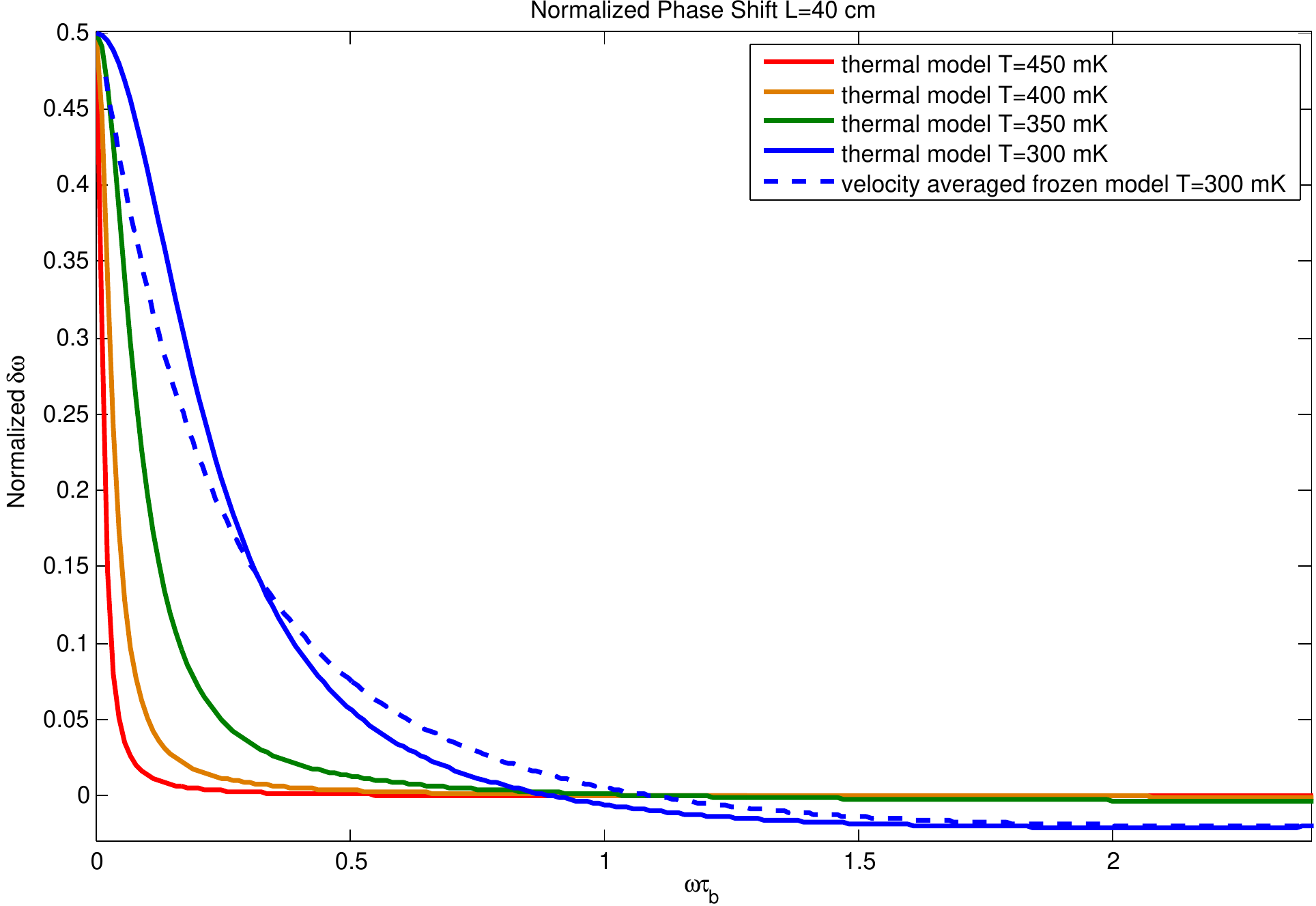}
\end{center}
\caption{The normalized spectrum of the linear in E phase shift for dilute
$^{3}$He dissolved in superfluid $^{4}$He, with temperature as a parameter.
All the solid curves are derived from the thermalization model, the dashed
line is the velocity averaged frozen model \cite{barabanov}.}%
\label{fig:thermshift}%
\end{figure}

\begin{figure}[ptbptb]
\begin{center}
\includegraphics[width=0.7\textwidth]{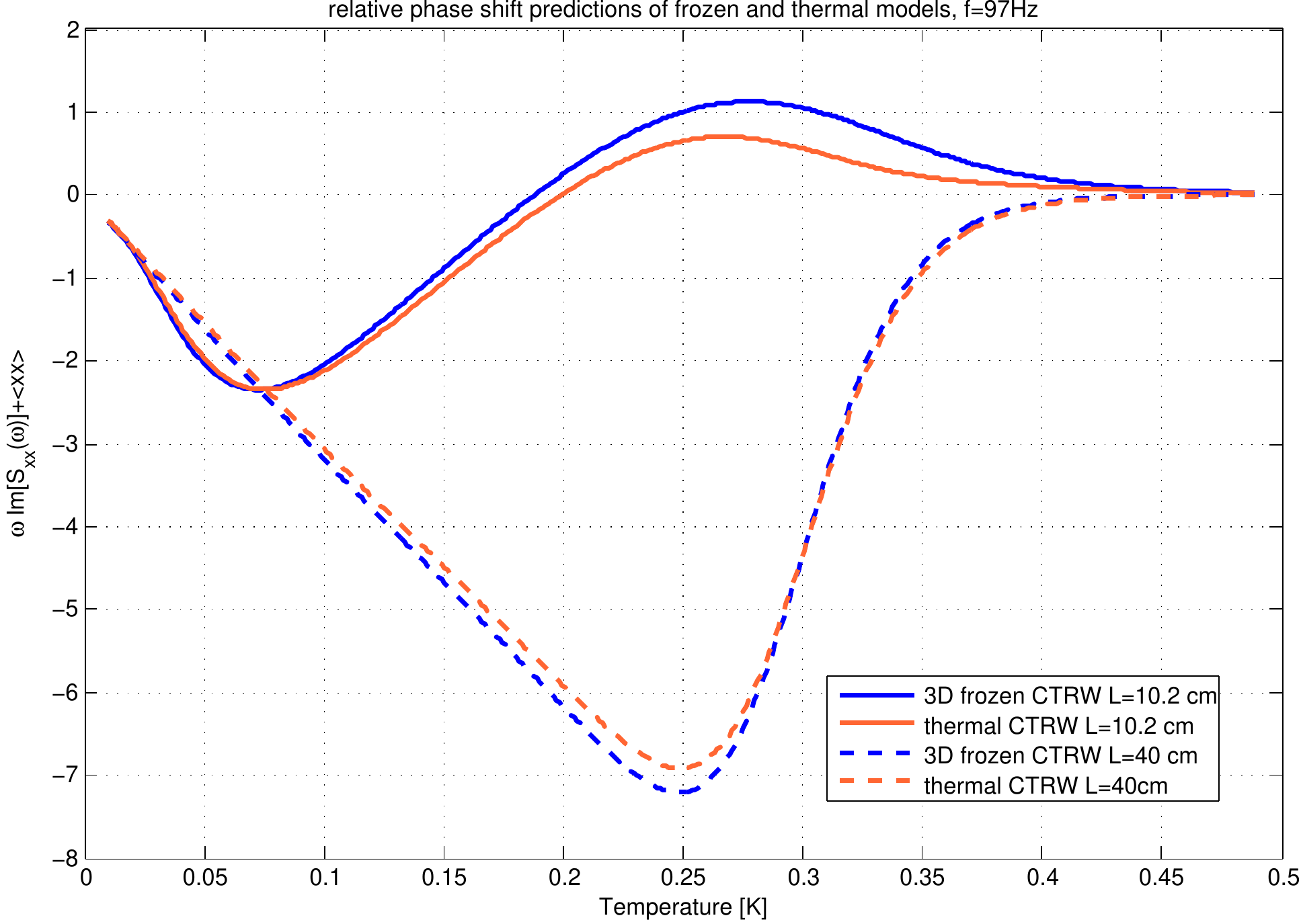}
\end{center}
\caption{The relative phase shift predicted from the thermalization model is
compared to the single velocity models averaged over velocity. As the
temperature decreases the models diverge, only the thermalization model
remains accurate as temperature is decreased. }%
\label{fig:thermshiftT}%
\end{figure}

For a number density ratio $^{3}$He:$^{4}$He $<$ $10^{-7},~^{3}$He-$^{3}$%
He$~$collisions can be ignored and collisions with the excitations in the
superfluid dominate. The system is taken to be a rectangle of 10.2 by 40 by
7.6 cm. In superfluid helium viscosity is absent $\cite{wilks}$ and the $^{3}%
$He behaves as if it were in a vacuum with an increased mass $m_{^{3}He}%
^{\ast}=2.4m_{^{3}He}$.~The $^{3}$He will thermalize by scattering on the
excitations, phonons and rotons, in the superfluid. When the temperature of
the superfluid is brought below $500~mK,$ phonons become the dominant
excitation. In such a system the diffusion coefficient was measured
$\cite{helium3diffusion, beck2015}$ and the data was fit well by the
equation,
\begin{align}
D=\frac{1.6}{T^{7}}.
\end{align}
We convert this to a collision time according to equation~(\ref{eq:tauctodiff}),%

\[
\tau_{c}=1.6\frac{m}{kT^{8}}.
\]
The predicted result is shown in figure~\ref{fig:thermshift} and as a function
of temperature in figure~\ref{fig:thermshiftT} along with the result from
\cite{barabanov}. The treatment of temperature is different in reference
\cite{barabanov}, where the single velocity random walk result is averaged
over a Maxwellian distribution of velocities. However an important prediction
remains; a strong dependence on temperature of the magnitude of the linear in
E shift. Therefore varying the temperature is a tool to mitigate and study the effect.



\subsubsection{ Comparison of the CTRWT model with Monte-Carlo simulations}

\begin{figure}[ptb]
\begin{center}
\includegraphics[width=0.85  \textwidth
]{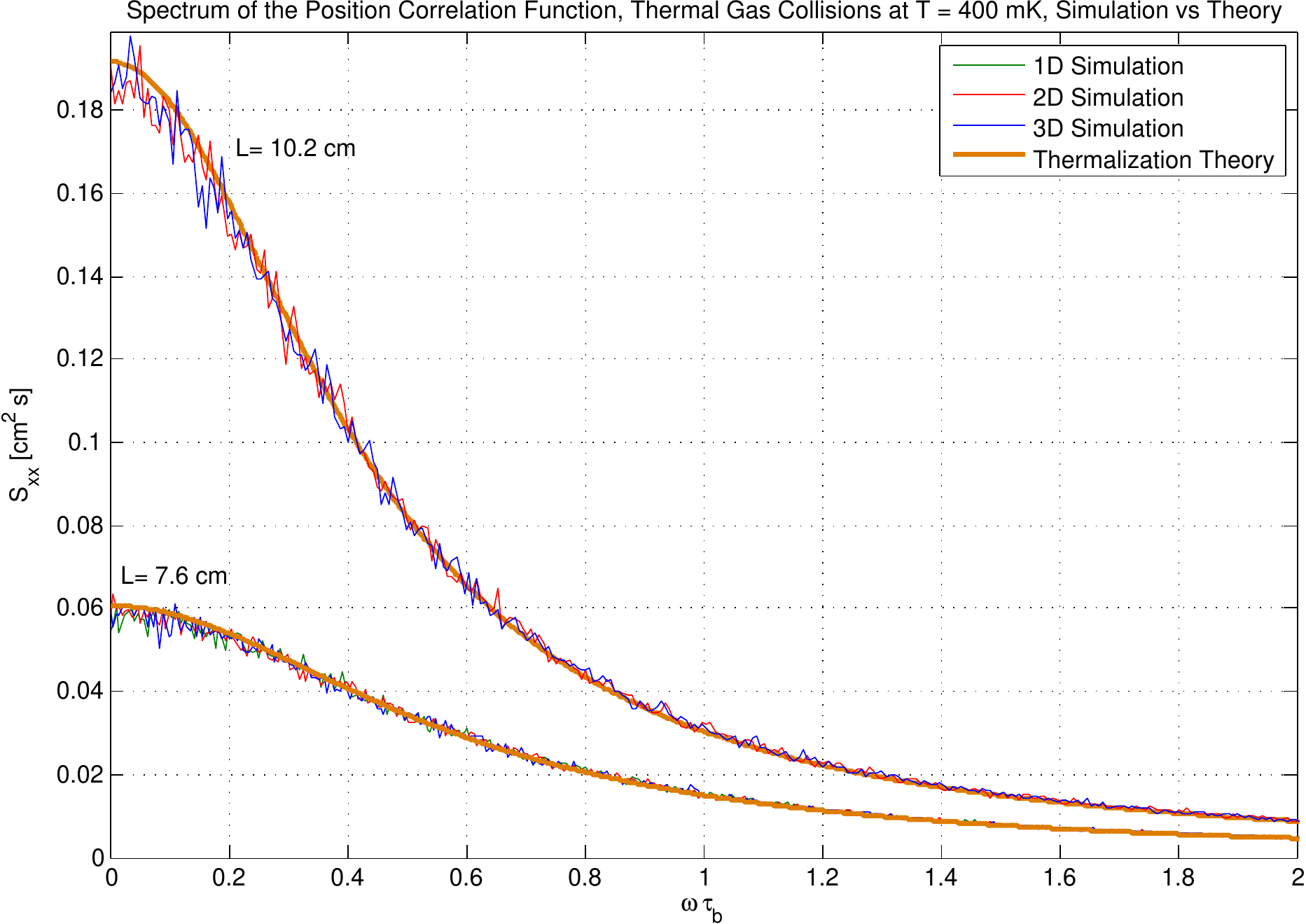}
\end{center}
\caption{The spectrum of the position autocorrelation function, a comparison
of theory to 1D, 2D and 3D simulations with thermalizing collisions in the
bulk. Good agreement is observed. Plotted are the correlation functions for
one direction in which the cell length is either 7.6 cm or 10.2 cm. The third
dimension in the 3D simulation is 40 cm, it is not shown. }%
\label{fig:thermspec}%
\end{figure}

\begin{figure}[ptb]
\begin{center}
\includegraphics[width=0.85\textwidth]{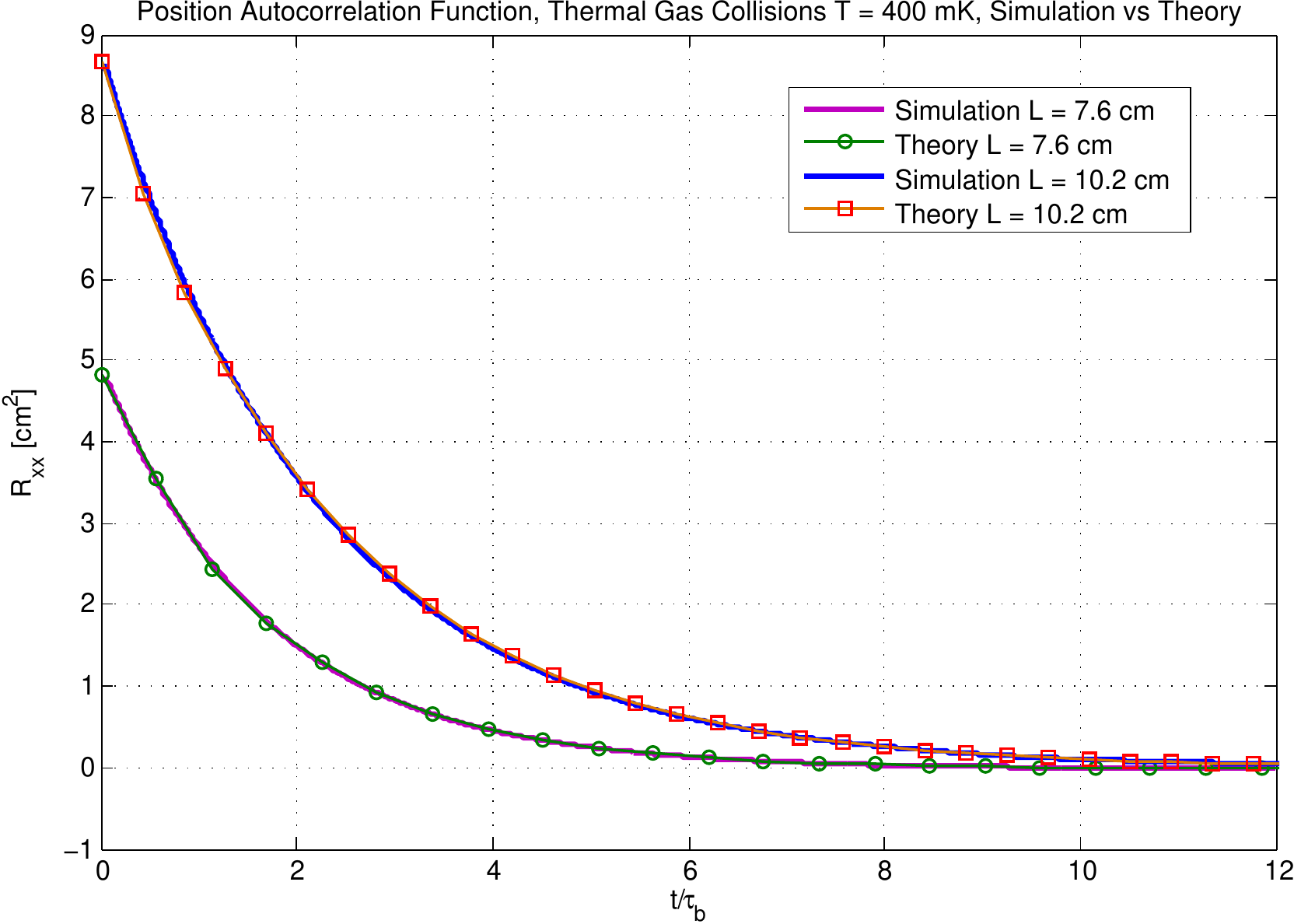}
\end{center}
\caption{The position autocorrelation function. The 3D Simulation is compared
to thermalization theory. Good agreement is observed.}%
\label{fig:thermcorr}%
\end{figure}

A comparison with 1D 2D and 3D Monte Carlo simulations are done on $10^{3}$
trajectories for $2\times10^{6}$ time steps. The trajectories are specific to
$^{3}$He at very low concentrations in superfluid $^{4}$He at 400 mK,
described in the previous section~\ref{sec:bounddomain}. In this regime the
mean free path is determined by collisions with phonons in the superfluid.
Upon a collision the new velocity was determined according to the isotropic 3D
Maxwellian distribution. The trajectories are confined by specular wall
collisions inside a rectangular volume 10.2 by 7.6 by 40 cm. The theoretical
spectrum of the position autocorrelation function (\ref{eq:BG3}) is shown in
figure~\ref{fig:thermspec}, and compared to the results of the simulations.
Figure~\ref{fig:thermcorr} shows the position autocorrelation function, a
function of time. The theoretical value of the position autocorrelation
function is found from numerical inversion of the theoretical result for the
spectrum except at $t=0$. Due to the finite nature of the numerical inversion
the $t=0$ point is obtained by the mean squared average of position,
$\left\langle x(t)x(t)\right\rangle $.

\bigskip

\section{Conclusion}

We have constructed a microscopic theory of the propagator (conditional
probability density) for a persistent random walk where the particles undergo
either Markovian stochastic scattering events or $1/v$ scattering satisfying
detailed balance and maintaining thermal equilibrium in both cases. For a gas
with a Maxwell-Boltzmann velocity distribution we obtain a relatively simple
expression for the propagator. The result is independent of the number of
dimensions considered, contrary to the ''frozen'' walk (a CTRW with fixed
velocity) where the number of dimensions in the walk strongly effect the
resonant structure of the correlation functions generated by the walk
$\cite{swank}$, and valid for \ all values of the scattering mean free path
from the quasi-ballistic to the diffusion regime of motion. We have shown
directly that our results go over into the standard diffusion theory for short
collision times (short mean free paths). We have shown how the results can be
applied to bounded regions using the method of images and have given results
for the position-position, postion-velocity and velocity-velocity correlation
functions, all of which have direct applications in calculating frequency
shifts and relaxation rates in nmr systems. One application is to the
calculation of the NMR phase shift of $^{3}$He in superfluid $^{4}$He in a
magnetic and electric field. The results differ somewhat from the previous
results obtained by averaging the 'frozen' walk results over a Maxwell
velocity distribution.

The method can be applied to inhomogeneous fields of any shape. We have
discovered a universal long-time tail $\propto t^{-1/2}$ independent of
dimensionality in bound systems. We emphasize that this long-time tail is expected only for bounded systems and it is diminished
with the increase of the system size. While we show that this effect is predicted
by the standard diffusion theory in agreement with \cite{oppenheim1964} who
found a similar long-time tail by solving the one dimensional Langevin
equation, the  independence of dimensionality does not seem to have been
noticed before.

\section{Acknowledgments}
We are grateful for fruitful discussion with Efim Katz on the long-time tail problem and with Bart McGuyer concerning scattering kernels. This work was supported in part by by the US Department of Energy under Grant No. DE-FG02-97ER41042.
\appendix

\section{Appendix}

\subsection{Calculation of the long-time tail from the thermalization model
\label{longtimetailderiv}}

It is useful to introduce dimension-less variables,%

\begin{equation}
\xi= \frac{\tau_{c}}{\tau_{b}}\text{ }\text{and}\text{ }\omega^{\prime}=
\omega\tau_{c}.%
\end{equation}

In terms of new variables equations (\ref{12}, \ref{13}) yield%

\begin{align}
z_{n}  &  =\frac{1+i\omega^{\prime}}{\sqrt{2}\xi\pi n},\\
S_{xx}\left(  \omega^{\prime}\right)   &  =\frac{16\tau_{c}L_{x}{}^{2}}{\pi
^{4}}\left(  -\frac{\pi^{4}}{96}+\sum_{n=1,3,\text{...}}^{\infty}\frac
{1}{n^{4}}\text{Re}\left[  1\left/  \left(  1-\left(  \sqrt{2\pi}n\xi\right)
^{-1}\text{e}^{z_{n}^{2}}\text{erfc}\left(  z_{n}\right)  \right)  \right.
\right]  \right) . \label{as2}%
\end{align}

The diffusive regime of motion is defined by $\xi<<1,$thus for not too high n,
$z_{n}$ $>>$1. Due to the strong cut-off by the prefactor $n^{-4}$ in
(\ref{as2}) only a few lower order terms are effective, which allows us to
apply the asymptotic expansion for $\operatorname*{erfc}\left(  z\right)  $
valid for $z>>1$, \cite{abramowitz} 7.1.23,%

\begin{equation}
\text{e}^{z^{2}}\text{erfc}\left(  z\right)  \approx\frac{1}{\sqrt{\pi}%
z}\left(  1-\frac{1}{2z^{2}}\right).  \label{as1}%
\end{equation}

Using (\ref{as1}) we can write for the sum in (\ref{as2}):%

\begin{align}
&  \sum_{n=1,3,\text{...}}^{\infty}\frac{1}{n^{4}}\text{Re}\left[  1\left/
\left(  1-\frac{1}{\sqrt{2\pi}n\xi}\frac{1}{\sqrt{\pi}z_{n}}\left(  1-\frac
{1}{2z_{n}{}^{2}}\right)  \right)  \right.  \right] \\
&  =\text{Re}\left[  \sum_{n=1,3,\text{...}}^{\infty}\frac{1}{n^{4}}%
\frac{\left(  -i+\omega^{\prime}\right)  ^{3}}{in^{2}\pi^{2}\xi^{2}%
+\omega^{\prime}\left(  -i+\omega^{\prime}\right)  ^{2}}\right]
=\text{Re}\left[  \sum_{n=1,3,\text{...}}^{\infty}\frac{1}{n^{4}}\frac{\left(
-i+\omega^{\prime}\right)  /\omega^{\prime}}{1+\alpha^{2}n^{2}}\right],
\label{as3}%
\end{align}

where,%

\begin{equation}
\alpha^{2}=\frac{i\pi^{2}\xi^{2}}{\omega^{\prime}\left(  -i+\omega^{\prime
}\right)  ^{2}} \label{as4}.%
\end{equation}

The sum in (\ref{as3}) converges,

\begin{equation}
S_{n}\left(  \omega^{\prime}\right)  =\sum_{n=1,3,\text{...}}^{\infty}\frac
{1}{n^{4}}\frac{\left(  -i+\omega^{\prime}\right)  /\omega^{\prime}}%
{1-\alpha^{2}n^{2}}=\frac{\left(  -i+\omega^{\prime}\right)  /\omega^{\prime}%
}{96}\pi\left(  \pi^{3}-12\pi\alpha^{2}+24\alpha^{3}\text{tanh}\left(
\frac{\pi}{2\alpha}\right)  \right).
\end{equation}

replacing $\alpha$ by (\ref{as4}),%

\begin{equation}
\operatorname{Re}[S_{n}\left(  \omega^{\prime}\right)  ]=\frac{\pi^{4}}%
{96}+\frac{\pi^{4}\xi^{2}}{8\omega^{\prime}{}^{2}\left(  1+\omega^{\prime}%
{}^{2}\right)  }-\text{Re}\left[  \frac{(-1)^{3/4}\pi^{4}\xi^{3}%
\text{tanh}\left(  \frac{(-1)^{3/4}\sqrt{\omega^{\prime}}\left(
-i+\omega^{\prime}\right)  }{2\xi}\right)  }{4\left(  \omega^{\prime}\right)
^{5/2}\left(  -i+\omega^{\prime}\right)  ^{2}}\right].  \label{as5}%
\end{equation}

Expanding and taking the real part in (\ref{as5}) we arrive at,%

\begin{equation}
Re[S_{n}(\omega^{\prime})]=\frac{\pi^{4}}{96}+\frac{\pi^{4}\xi^{2}}{8\left(
\omega^{\prime}\right)  ^{2}\left(  1+\omega^{\prime}{}^{2}\right)  }\left(
1-\Delta\left[  \xi,\omega^{\prime}\right]  \right).  \label{as6}%
\end{equation}
with $\Delta\left[  \xi,\omega^{\prime}\right]  $ given by (\ref{as7}).

\subsection{Calculation of the long time tail in ordinary diffusion theory}

We start from the well known relation,%

\begin{equation}
R_{v_{x}v_{x}}(t)=-\partial_{t}^{2}R_{xx}(t). \label{1}%
\end{equation}

For a 3D diffusive motion in a rectangular domain the auto-correlation
function of the displacement in each direction is given by a term of the form,%

\begin{equation}
R_{xx}(t)=\frac{8L_{x}{}^{2}}{\pi^{4}}\sum_{n=0}^{\infty}\frac{1}{(2n+1)^{4}%
}e^{-\frac{(2n+1)^{2}t}{\tau_{d}}} .\label{2}%
\end{equation}

Where,%

\begin{equation}
\tau_{d}=\frac{L_{x}{}^{2}}{\pi^{2}D},%
\end{equation}

is the time constant for the lowest diffusion mode. Expression (\ref{2}) is
valid for not too short times, $t>>\tau_{c}$, $\left(  \tau_{c}\text{ is the
}\text{mean }\text{time }\text{between }\text{particle }\text{collisions}%
\right)  $. Inserting (\ref{2}) into (\ref{1}) we find, in agreement with (62)
in reference \cite{oppenheim1964},%

\begin{equation}
R_{v_{x}v_{x}}(t)=-\frac{8L_{x}{}^{2}}{\pi^{4}}\frac{1}{\tau_{d}{}^{2}}%
\sum_{n=0}^{\infty}e^{-\frac{(2n+1)^{2}t}{\tau_{d}}}=-\frac{4L_{x}{}^{2}}%
{\pi^{4}}\frac{1}{\tau_{d}{}^{2}}\vartheta_{2}\left(  0,e^{-\frac{4t}{\tau
_{d}}}\right),  \label{4}%
\end{equation}

where $\vartheta_{2}(u,z)$ is Jacobi theta function.

To investigate the short-time, ( $t$ $<<\tau_{d}$), behavior of the Jacobi
theta function we expand it in a series and keep only lowest order terms,%

\begin{equation}
\vartheta_{2}\left(  0,e^{-\frac{4t}{\tau_{d}}}\right)  \approx\frac{\sqrt
{\pi}\sqrt{\tau_{d}}}{2\sqrt{t}}. \label{5}%
\end{equation}

Inserting (\ref{5}) into (\ref{4}) we find,%

\begin{equation}
R_{v_{x}v_{x}}(t)=-\frac{4L^{2}}{\pi^{4}}\frac{1}{\tau_{d}^{2}}\frac{\sqrt
{\pi}\sqrt{\tau_{d}}}{2\sqrt{t}}=-\frac{2}{\pi^{1/2}}\xi\frac{kT}{m}\left(
\frac{t}{\tau_{c}}\right)^{-1/2} .\label{6}%
\end{equation}

For a longer times, ($t$ $>>$$\tau_{d}$ ), $e^{-\frac{4t}{\tau_{d}}}%
$$\rightarrow$0 and we may expand $\vartheta_{2}(0,z)$ for z$\rightarrow$0.
Again keeping only lowest order terms,%

\begin{equation}
R_{v_{x}v_{x}}(t)=-\frac{8L^{2}}{\pi^{4}}\frac{1}{\tau_{d}^{2}}\text{exp}%
\left[  -\frac{t}{\tau_{d}}\right] . \label{7}%
\end{equation}

We see that (\ref{6}) is exactly the same as second term in (\ref{as12}).
Hence, both Diffusion theory and our CTRWT model predicts the existence of a
long -time tail (\ref{6}) in the correlation function for each velocity
component. This negative tail exists for $\tau_{c}$$<<$ t $<<$$\tau_{d}$, for
even longer times the velocity correlation function decays exponentially with
time constant $\tau_{d}$, see (\ref{7}).

\section{\bigskip Cartesian Projection from 3D, equivalence for arbitrary
velocity distributions.\label{sec:1D2D3Dagree}}

\begin{figure}[ptb]
\begin{center}
\includegraphics[width=0.4\textwidth]{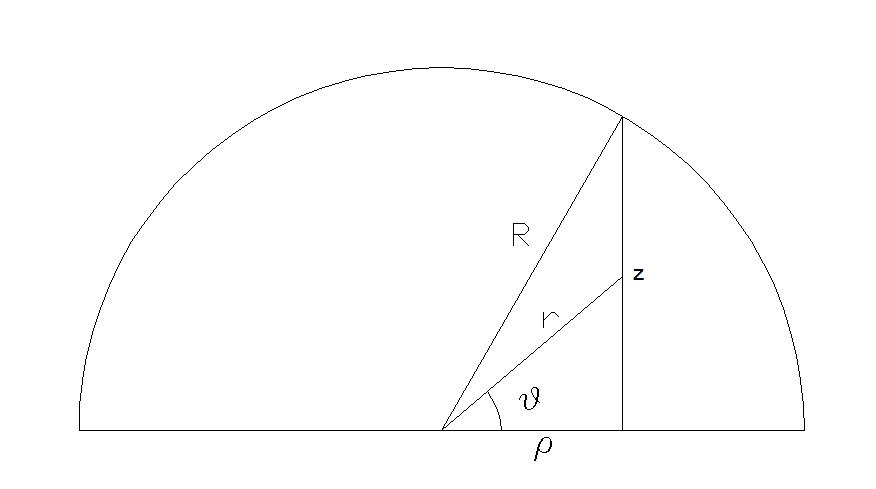}
\end{center}
\caption{Visualization of the projection from a higher dimension onto a lower
dimension.}%
\label{fig:proj}%
\end{figure}\begin{figure}[ptbptb]
\begin{center}
\includegraphics[width=0.4\textwidth]{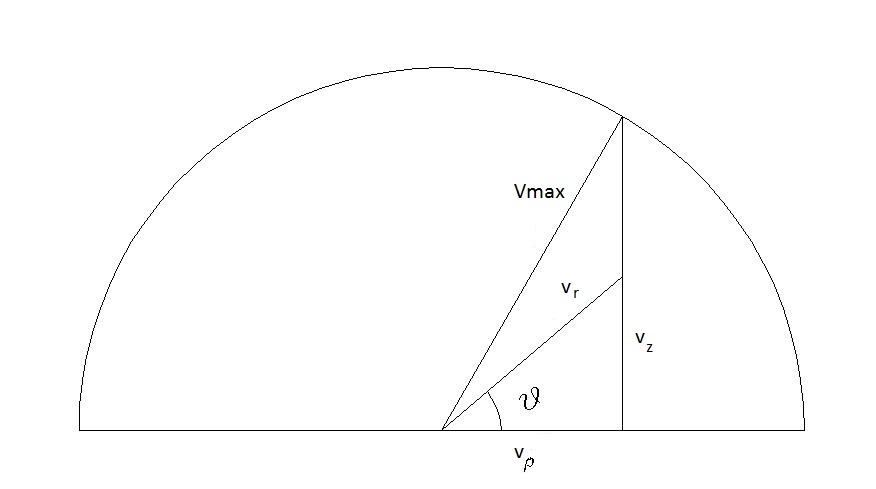}
\end{center}
\caption{Visualization of the projection from a higher dimension onto a lower
dimension.}%
\label{fig:vproj}%
\end{figure}The CTRW probability density defines how trajectories propagate
and is typically used to calculate averages and correlation functions. In the
case that the CTRW probability density function is isotropic, and the velocity
distribution associated with the CTRW is isotropic, and the function that is
being averaged or correlated does not depend on one or more of the Cartesian
coordinates, then the variable can be integrated away. The resulting function
will only depend on one or two of the Cartesian coordinates. This is because
cross-correlation between the different coordinates in a Cartesian system is
absent. The CTRWT can be expressed in terms of the spectrum of the $F\left(
\mathbf{r},t\right)  ,$ function as given by the equation,
\begin{equation}
p(\mathbf{q},\omega\mathbf{)}=\frac{F(\mathbf{q},\omega)}{1-\frac{1}{\tau_{c}%
}F(\mathbf{q},\omega)},
\end{equation}
where,
\begin{equation}
F_{ND}(\mathbf{x},t)=\int\alpha_{ND}(\mathbf{v})G_{ND}(\mathbf{x,}t,\mathbf{v}%
)d^{N}\mathbf{v}.
\end{equation}

We wish to verify that,
\begin{equation}
F_{2D}(\mathbf{x},t)=\int F_{3D}(\mathbf{x},t)dz. \label{eq:goal}%
\end{equation}

During the remainder of this proof the 2D side is on the left, and the 3D side
is on the right. Writing this in terms of the velocity averaged $G$
functions,
\begin{align}
\int\alpha_{2D}(v)G_{2D}(\mathbf{x,}t,\mathbf{v})d^{2}\mathbf{v}  &  =\int
\int\alpha_{3D} (v)_{3D}G_{3D}(\mathbf{x,}t,\mathbf{v})d^{3}\mathbf{v}dz,\\
\int\alpha_{2D}(v)\delta^{2}(\mathbf{x}-\mathbf{v}t)e^{-\frac{t}{\tau_{c} }%
}d^{2}\mathbf{v}  &  =\int\int\alpha_{3D} (v)\delta^{3}(\mathbf{x}-\mathbf{
v}t)e^{-\frac{t}{\tau_{c}}}d^{3}\mathbf{v}dz. \label{eq:cart1}%
\end{align}

\bigskip In polar coordinates we can define the delta function,%

\begin{equation}
\delta^{2}(\mathbf{x})=\frac{\delta(\rho)}{\rho}\delta(\phi),
\end{equation}

where $\rho=\sqrt{x^{2}+y^{2},}$ and $\phi=\arctan(y/x).$ Similarly in
spherical coordinates we have,%

\begin{equation}
\delta^{3}(\mathbf{x})=\frac{\delta(r)}{r^{2}}\frac{\delta(\theta)}{\sin
\theta}\delta(\phi),
\end{equation}
with, $r=\sqrt{x^{2}+y^{2}+z^{2}},~\theta=\arctan(\sqrt{x^{2}+y^{2}%
}/z)$ and $\phi=\arctan(y/x).$

Inserting these definitions into equation (\ref{eq:cart1}) and writing the
integration over polar and spherical coordinates we have,
\begin{align}
&\int\int\alpha_{2D} (v)\frac{\delta(\rho-vt)}{\rho}\delta(\phi-\phi
_{v})e^{-\frac{t}{\tau_{c}}}vdvd\phi_{v}\\
&=\int\int\alpha_{3D} (v)\frac{
\delta(r-vt)}{r^{2}}\frac{\delta(\theta-\theta_{v})}{\sin\theta}\delta
(\phi-\phi_{v})e^{-\frac{t}{\tau_{c}}}v^{2}\sin\theta_{v}dvd\theta_{v}%
d\phi_{v}dz.
\end{align}
Integration over the angular coordinates in each gives,
\begin{equation}
\int~\alpha_{2D} (v_{\rho})\frac{\delta(\rho-v_{\rho}t)}{\rho}e^{- \frac
{t}{\tau_{c}}}v_{\rho}dv_{\rho}=\int\int\alpha_{3D} (v)\frac{ \delta
(r-vt)}{r^{2}}e^{-\frac{t}{\tau_{c}}}v^{2}dvdz. \label{eq:cart2}%
\end{equation}

We can continue if we assume that $z$ can be determined from $r$ and $\rho$
according to figure \ref{fig:proj} by,
\begin{equation}
z=\sqrt{r^{2}-\rho^{2}}.
\end{equation}
Therefore if we wish to integrate over $z,$ the definition of a projection, we
can change this to in integration over $r$ by,
\begin{equation}
dz=\frac{2rdr}{\sqrt{r^{2}-\rho^{2}}}.
\end{equation}

Inserting this into the right hand side in equation (\ref{eq:cart2}),%

\begin{align}
\int~\alpha_{2D} (v_{\rho})\frac{\delta(\rho-v_{\rho}t)}{\rho}e^{- \frac
{t}{\tau_{c}}}v_{\rho}dv_{\rho}  &  =\int\int_{\rho}^{\infty}\alpha
_{3D}(v)\frac{\delta(r-vt)}{r^{2}}e^{-\frac{t}{\tau_{c}}}v^{2}dv\frac
{2rdr}{\sqrt{r^{2}-\rho^{2}}},\\
\int~\alpha_{2D} (v_{\rho})\frac{\delta(\rho-v_{\rho}t)}{\rho}e^{- \frac
{t}{\tau_{c}}}v_{\rho}dv_{\rho}  &  =\int_{0}^{\infty}\alpha_{3D}%
(v)e^{-\frac{t}{\tau_{c}}}\frac{2vdv}{t\sqrt{vt^{2}-\rho^{2}}}\left(
1-\Theta(\rho-vt)\right).
\end{align}

\bigskip$\Theta(\rho-vt)$ is the Heaviside step function, it is needed because
$r\geq\rho$, and prevents the function from going imaginary. To continue we
define a $2D$ velocity distribution, $\alpha(v_{\rho})_{2D},$ as a projection
from the$~3D$ velocity distribution according to figure \ref{fig:vproj}. Let
us define the 2D velocity distribution as,%

\begin{equation}
\alpha_{2D}(v_{\rho})=\int_{v_{\rho}}^{\infty}\frac{\alpha_{3D}(v)2vdv}%
{\sqrt{v^{2}-v_{\rho}^{2}}}.%
\end{equation}

Inserting this we have,
\begin{equation}
\int~\int_{v_{\rho}}^{\infty}\frac{\alpha_{3D}(v)2vdv}{\sqrt{ v^{2}-v_{\rho
}^{2}}}\frac{\delta(\rho-v_{\rho}t)}{\rho}e^{-\frac{t}{ \tau_{c}}}v_{\rho
}dv_{\rho}=\int_{0}^{\infty}\alpha_{3D} (v)e^{-\frac{t}{ \tau_{c}}}\frac
{2vdv}{t\sqrt{(vt)^{2}-\rho^{2}}}\left(  1-\Theta(\rho-vt)\right).
\end{equation}
We continue by scaling the delta function,
\begin{equation}
\int\int_{v_{\rho}}^{\infty}\frac{\alpha_{3D}(v)2vdv}{\sqrt{v^{2}-v_{\rho}%
^{2}}}\frac{\delta(v_{\rho}-\frac{\rho}{t})}{t\rho}e^{-\frac{t}{\tau_{c}}%
}v_{\rho}dv_{\rho}=\int_{0}^{\infty}\alpha_{3D}(v)e^{-\frac{t}{\tau_{c}}}%
\frac{2vdv}{t\sqrt{(vt)^{2}-\rho^{2}}}\left(  1-\Theta(\rho-vt)\right).
\end{equation}

Integrating over $v_{\rho}$ we find,%
\begin{align}
\int_{\frac{\rho}{t}}^{\infty}\frac{\alpha_{3D}(v)2vdv}{\sqrt{ v^{2}-\left(
\frac{\rho}{t}\right)  ^{2}}}\frac{1}{t\rho}e^{-\frac{t}{\tau_{c}}}\frac{\rho
}{t} \left(  1-\Theta(-\frac{\rho}{t})\right)   &  =\int_{0}^{\infty}%
\alpha_{3D}(v)e^{-\frac{t}{\tau_{c}}}\frac{2vdv}{t\sqrt{(vt)^{2}-\rho^{2}}%
}\left(  1-\Theta(\rho-vt)\right), \\
\int_{\frac{\rho}{t}}^{\infty}\frac{\alpha_{3D}(v)2vdv}{t\sqrt{ (vt)^{2}%
-\rho^{2}}}e^{-\frac{t}{\tau_{c}}} \left(  1-\Theta(-\frac{\rho}{t})\right)
&  =\int_{0}^{\infty}\alpha_{3D}(v)e^{-\frac{t}{\tau_{c}}}\frac{2vdv}%
{t\sqrt{(vt)^{2}-\rho^{2}}}\left(  1-\Theta(\rho-vt)\right).
\end{align}

Again the Heaviside is there to keep the function real. In general we have
$v>0,$ and $r\geq\rho$, so we can take the real part of the integral from zero
to infinity,%

\begin{equation}
\mathrm{Re}\left[ \int_{0}^{\infty}\frac{\alpha_{3D}(v)2vdv}{t\sqrt
{(vt)^{2}-\rho^{2}}}e^{- \frac{t}{\tau_{c}}}\right]  =\mathrm{Re}\left[
\int_{0}^{\infty}\frac{\alpha_{3D}(v)2vdv}{t\sqrt{(vt)^{2}-\rho^{2}}}%
e^{-\frac{t}{\tau_{c}}}\right] .
\end{equation}

We have verified that equation (\ref{eq:goal}) is valid. Therefore with an
isotropic velocity distribution we can find the results of the projected $3D$
random walk with the projected velocity distribution and the $2D$ random walk.
We should point out that this projection is already satisfied by Maxwellian
distributions, where we found that the conditional density is the same for
$1D,$ $2D$ and $3D.$

The spectrum of a $3D$ CTRWT is a projection if the function in question does
not depend on one or more of the Cartesian coordinates of the $3D$ system.
Regardless of dimensions used in the model if a function does not depend on a
particular Cartesian coordinate that coordinate can be integrated away.
Consider the spectrum of an arbitrary function $h(\rho,\phi,z)=h(\rho,\phi)$
in cylindrical coordinates,
\begin{align}
\int_{-\infty}^{\infty}h(\rho,\phi)e^{-i\mathbf{q\cdot x}}d^{3}\mathbf{x}%
=H(q_{\rho},q_{\phi})\delta(q_{z}).
\end{align}
We have $q_{z}=0$ for all $z$. This is equivalent to integrating over the $z$
direction, and is the definition we used as a projection onto the $x,y$ plane.
Thus, when we solve for the Fourier transform we are automatically taking the
projection onto the plane normal to the $z$ direction. Therefore isotropic
velocity distributions in Cartesian coordinates allows the random walk in
$1D~$or $2D$ to solve for the projections of the $3D$ random walk.

\bibliographystyle{ieeetr}
\bibliography{SwankBibliography}

\bigskip
\end{document}